\documentclass[onecolumn]{emulateapj}
\usepackage{graphicx}
\usepackage{longtable}

\def\er{$\pm$}

\begin{document}

\title{Systematic Uncertainties in the Spectroscopic Measurements of
  Neutron-Star Masses and Radii from Thermonuclear X-ray
  Bursts. II. Eddington Limit}

\author{Tolga   G\"uver\altaffilmark{\dag},  Feryal   \"Ozel,   \&  Dimitrios
  Psaltis}

\affil{Department of  Astronomy, University of Arizona,  933 N. Cherry
  Ave., Tucson, AZ 85721} 

\altaffiltext{\dag}{Current Address : Sabanc\i ~University, Faculty
  of  Engineering and  Natural Sciences,  Orhanl\i ~- Tuzla,  Istanbul
  34956, Turkey}

\begin{abstract}
  Time  resolved X-ray spectroscopy  of thermonuclear  bursts observed
  from low mass X-ray binaries  offer a unique tool to measure neutron
  star masses  and radii.  In  this paper, we continue  our systematic
  analysis  of all  the X-ray  bursts  observed with  RXTE from  X-ray
  binaries.   We determine the  events which  show clear  evidence for
  photospheric radius  expansion and measure the  Eddington limits for
  these accreting  neutron stars using the  bolometric fluxes attained
  at the touchdown  moments of each X-ray burst.  We employ a Bayesian
  technique to investigate the degree to which the Eddington limit for
  each  source  remains constant  between  bursts.  We  find that  for
  sources with  a large number of radius  expansion bursts, systematic
  uncertainties  are at a  5$-$10\% level.   Moreover, in  six sources
  with  only pairs  of Eddington-limited  bursts, the  distribution of
  fluxes is consistent with a $\sim 10$\% fractional dispersion.  This
  indicates that the spectroscopic measurements of neutron star masses
  and radii  using thermonuclear X-ray  bursts can reach the  level of
  accuracy  required  to distinguish  between  different neutron  star
  equations  of  state, provided  that  uncertainties  related to  the
  overall  flux  calibration  of  X-ray detectors  are  of  comparable
  magnitude.
\end{abstract}

\keywords{stars: neutron --- X-rays: bursts}

\section{Introduction}

Measurements of the masses and  radii of neutron stars provide some of
the most direct constraints on the  equation of state of the matter in
the cores  of these compact objects. Time  resolved X-ray spectroscopy
of  thermonuclear bursts  observed from  some  of the  low mass  X-ray
binaries has  been one of  the observational methods to  constrain the
neutron star  masses and radii  (see, e.g., van Paradijs\  1978, 1979;
Damen et al.\  1990; Lewin, van Paradijs, \&  Taam\ 1993).  The method
involves  modeling high  time resolution,  high  signal-to-noise X-ray
burst data  to spectroscopically measure  the apparent radius  and the
Eddington luminosity for  the neutron star, both of  which are related
to its mass and radius.

The first  few seconds of  some of the  brightest X-ray bursts  show a
characteristic pattern  in which  the color temperature  increases and
then  decreases,  while the  apparent  radius monotonically  increases
(see, e.g.,  Galloway et al.\ 2008a). Eventually,  the apparent radius
starts to  decrease as  the color temperature  reaches a peak  and the
burst  starts to  decay.  In  the  meantime, the  flux remains  nearly
constant  at a  peak  value.   This phenomenon  is  understood as  the
response  of   the  outermost  layers   of  the  neutron  star   to  a
super-Eddington burst flux, where the photosphere expands to few times
the stellar radius and subsequently contracts back to the neutron star
surface.  During  the expansion and  the contraction phase,  the X-ray
flux stays very close to the  Eddington limit and any excess energy is
transferred into kinetic energy of  the outflow (see, e.g., Kato 1983;
Ebisuzaki, Hanawa,  \& Sugimoto 1983; Paczynski  \& Proszynski\ 1986).
Accordingly, X-ray  bursts from which this phenomenon  is observed are
called  Photospheric  Radius Expansion  (PRE)  events  and the  fluxes
attained during the  expansion episodes of these bursts  are used as a
measure  of  the  local  Eddington  limit of  the  neutron  star  (van
Paradijs\  1978), where  the  gravitational and  radiation forces  are
balanced.

PRE events  can be used to  determine the Eddington  luminosity if the
distance  to  the  X-ray  binary  is known  (see,  e.g.,  Basinska  et
al.\  1984, Damen et  al.\ 1990;  Kuulkers et  al.\ 2003,  Galloway et
al.\ 2008a, b). Using the  X-ray bursters located in globular clusters
and  the  peak  fluxes   reached  during  X-ray  bursts,  Kuulkers  et
al.\  (2003) tested the  idea that  the PRE  events can  be used  as a
standard candle.  They found that  the peak fluxes attained during the
PRE events can indeed be used  as standard candles and are accurate to
at least  within 15\%.  Similarly, using  66 and 40  X-ray bursts from
4U~1728$-$34 and  4U~1636$-$536, Galloway  et al.\ (2003,  2006) found
that  the peak  fluxes  reached during  photospheric radius  expansion
events  are   normally  distributed  with  a   standard  deviation  of
$\approx$~3\%  and 7.6\%, respectively,  after corrections  related to
the  orbital modulation  and  the composition  of  the atmosphere  are
applied.

Even  though a  measurement of  the  Eddington limit  of an  accreting
neutron  star is  useful toward  measuring  its mass  and radius,  the
determination of  the exact  moment when a  given X-ray  burst reaches
this  limit is not  always straightforward.   The observed  X-ray flux
during the  photospheric radius expansion episode is  expected to vary
due  to  changes  in   the  gravitational  redshift  as  the  apparent
photospheric radius rises  and falls (see, e.g., Damen  et al.\ 1990).
The first  moment the flux  reaches the Eddington limit  occurs during
the burst  rise and is not  always robustly identified for  all of the
bursts.   Alternatively, the Eddington  limit can  be measured  at the
moment  when  the  photosphere  ``falls''  back to  the  neutron  star
surface.  This  has been  called the touchdown  moment (Damen  et al.\
1990)  and  is  identified  as   the  point  at  which  the  blackbody
temperature  reaches the  highest  value during  the  burst while  the
apparent radius is lowest. Combined with a measurement of the distance
and apparent angular size of  the neutron star, the measurement of the
Eddington flux  at touchdown can lead to  uncorrelated measurements of
the neutron star mass and radius (see, e.g., Ebisuzaki\ 1987; Damen et
al.\ 1990; \"Ozel et al.\ 2009; G\"uver et al.\ 2010a, b).

Nearly  continuous observations  of bursting  low mass  X-ray binaries
over the  last 15  years with the  Rossi X-ray Timing  Explorer (RXTE)
provided high  quality data  for over one  thousand X-ray  bursts from
more than  forty X-ray binaries  (Galloway et al.\ 2008a).   This rich
database of X-ray burst observations enables a study of the spectra of
PRE  bursts from which  the Eddington  limit can  be measured  and any
systematic variations in the inferred spectral parameters of the X-ray
bursts  can be  inferred. Such  an assessment  is essential  to better
establish  the reliability of  the mass  and radius  measurements from
time-resolved spectroscopic analysis of X-ray bursts.

Using  the  archival  RXTE   observations,  we  recently  studied  the
systematic uncertainties  present in the  apparent radius measurements
during the  cooling tails of the  X-ray bursts (G\"uver  et al.\ 2011,
hereafter Paper I). Our analysis  showed that the vast majority of the
X-ray spectra extracted from the cooling tails of 447 X-ray bursts are
statistically  consistent with  Planckian functions  and  the inferred
spectral parameters for the majority of the bursts follow the expected
F~$\propto$~T$^{4}$ relation  for most of the  sources.  These results
enabled us to measure the apparent  radii of a number of neutron stars
and assess the systematic uncertainties in these measurements.

In this paper, we continue to analyze all of the X-ray bursts observed
from low mass  X-ray binaries in order to  determine the uncertainties
related to  spectroscopic measurements of  the Eddington limit  in PRE
bursts.   We focus on  the measurement  of the  Eddington flux  at the
touchdown  moments in twelve  X-ray binaries  from which  multiple PRE
events  have been  observed. Our  aim is  to determine  any systematic
uncertainties in these measurements.

In  \S2,  we briefly  summarize  the  observations  and data  analysis
techniques, which  we discuss in  full detail in  Paper I. In  \S3, we
introduce a systematic method to  select the PRE events from the burst
archive using time resolved  spectroscopic measurements. In \S4 and 5,
we describe  the statistical tools based on  Bayesian Gaussian mixture
algorithms that we use to determine the Eddington limit and associated
systematic uncertainties for each source.  Finally, in \S6, we present
our results and discuss their implications.

\section{Observations and Data Analysis}

Galloway et al.\  (2008a) presented a catalog of  RXTE observations of
X-ray bursts  from 48 low mass  X-ray binaries. Following  Paper I, we
chose  12  X-ray  binaries from  this  sample  based  on a  number  of
criteria.  We  included only  the sources that  show at least  two PRE
events (as defined in Galloway  et al.\ 2008a).  We excluded all X-ray
binaries  that are  known to  be dippers,  ADC sources,  or  have high
inclinations as  well as the known millisecond  pulsars.  Because they
are likely to be affected by source confusion (Galloway et al.\ 2008a;
Keek et al.\ 2010),  we excluded observations of GRS~1741.9$-$2853 and
2E~1742.9$-$2929  and also a  small number  of bursts  from Aql~X$-$1,
4U~1728$-$34,  and 4U~1746$-$37.  Finally,  since a  study of  the PRE
events  observed from  EXO~1745$-$248, 4U~1608$-$52,  and 4U~1820$-$30
have been  reported elsewhere  (\"Ozel et al.   2009; G\"uver  et al.\
2010a, b), the results for these sources will not be repeated here.

As in  Paper I,  we imposed  a limit on  the persistent  flux measured
prior to  each X-ray burst  such that it  does not exceed 10\%  of the
peak burst  flux, i.e., $\gamma\equiv F_{\rm  per}/F_{\rm Edd}<0.1$ as
calculated by  Galloway et al.\  (2008a). Imposing this  limit reduces
the systematic  uncertainties introduced by  subtracting the pre-burst
emission from the X-ray burst spectra.

\begin{deluxetable}{ccccc}
  \tablecolumns{5}
  \tablewidth{370pt}
\tablecaption{THE NUMBER OF PRE EVENTS FOR EACH SOURCE}
  \tablehead{\colhead{Name} & \colhead{Number of
      Bursts\tablenotemark{a}}  & 
\colhead{Catalog PRE\tablenotemark{b}}  & \colhead{$\gamma$
  Limit\tablenotemark{c}} & 
\colhead{n$_{PRE}$}}
\startdata
4U~0513$-$40          &  7 & 2 & 2 & 2 \\
  4U~1636$-$53          &  172 & 52 & 49 & 46 \\
  4U~1702$-$429         &  47  & 6 & 6 & 1 \\
  4U~1705$-$44          &  47  & 4 & 4  & 2 \\
  4U~1724$-$307         &  3  & 3 & 3 & 2\tablenotemark{d} \\
  4U~1728$-$34          &  106 & 80 & 71 & 16 \\
  KS~1731$-$260         &  27  & 6 & 4 & 2 \\
  4U~1735$-$44          &  11  & 8 & 4 & 2 \\
  4U~1746$-$37          &  30 & 3 & 0 & -- \\
  SAX~J1748.9$-$2021    &  16 & 8 & 3 & 2 \\
  SAX~J1750.8$-$2900    &  4  & 2 & 2 & 2 \\
  Aql~X$-$1             &  57  & 10 & 10 & 6
\enddata
\tablenotetext{a} {Values  are adopted  from Galloway et  al.\ (2008a)
  and  show  the total  number  of  X-ray  bursts detected  by  RXTE.}
\tablenotetext{b} {The total  number of X-ray bursts tagged  as PRE or
  potentially  PRE events in  the Galloway  et al.\  (2008a) catalog.}

\tablenotetext{c}{The number  of remaining bursts with  peak flux that
  exceeds the pre-burst emission by a factor of 10.}

\tablenotetext{d}{As discussed  in detail in Paper I,  we excluded the
  first  burst observed  from 4U~1724$-$307  from our  analysis, since
  model fits of the X-ray spectra extracted from this burst can not be
  fitted with a Planckian function and addition of absorption edges at
  several energies is needed (in't Zand et al. 2010).}
\label{sourcestable}
\end{deluxetable}

The  final list  of all  the X-ray  binaries and  the X-ray  bursts we
studied  is presented in  Table~\ref{sourcestable}.  We  performed the
data  analysis   following  the   methods  detailed  in   Galloway  et
al.\   (2008a)  and   in  Paper   I.   We   extracted   time  resolved
2.5$-$25.0~keV X-ray spectra from  all the RXTE/PCA layers.  We varied
the exposure time  between 0.25~s and 1~s to  keep the signal-to-noise
ratio constant based on the count rate during the burst.  We also used
a  16~s  spectrum,  obtained  prior  to  each  burst,  as  background.
Response matrix  files were generated  using the PCARSP  version 11.7,
HEASOFT release  6.7, and  HEASARC's remote calibration  database.  We
took into account  the offset pointing of the  PCA during the creation
of the response matrix files.   Finally, we corrected all of the X-ray
spectra for  PCA deadtime following  the method suggested by  the RXTE
team.\footnote{ftp://legacy.gsfc.nasa.gov/xte/doc/cook\_book/pca\_deadtime.ps}

We used the Interactive Spectral Interpretation System (ISIS), version
1.4.9-55    (Houck    \&    Denicola    2000)   and    custom    built
S-Lang\footnote{http://www.jedsoft.org/slang/}  scripts  for  spectral
analysis.  We  fit each spectrum  with a blackbody function  using the
{\it bbodyrad} model (as defined  in XSPEC; Arnaud 1996) and with {\it
  tbabs}  (Wilms,  Allen,  McCray\  2000) to  model  the  interstellar
extinction.   For each source,  we fixed  the hydrogen  column density
(N$_{\rm H}$) to the  values given in Table 1 of Paper  I. In the same
analysis, we also determined  that the level of systematic uncertainty
required to  make the  X-ray burst spectra  of each  source consistent
with blackbody functions is less than 5\% (see Section 3.1 and Table 2
in Paper  I for  details).  During each  fit, we included  these minor
systematic uncertainties  that we inferred  for each source.   We then
created  for each  burst  that  has high  temporal  and spectral  data
coverage, a  time series of blackbody temperatures  $T_{\rm c}$ (units
of  keV)   and  normalizations  $A$  (in   units  of  [km/10~kpc]$^2$)
throughout  the burst  that resulted  from the  time-resolved spectral
analysis.   We  used Equation  (3)  of  Galloway  et al.   (2008a)  to
calculate the bolometric fluxes.  In the following sections we adopted
the burst numbering system introduced by Galloway et al. (2008a).

\section{Determination of Photospheric Radius Expansion Events}

Our first aim  is to select the PRE bursts in  the X-ray burst sample,
so that  we can use the  fluxes attained in  them as a measure  of the
local Eddington limit on the  neutron star surface.  As a signature of
PRE, we look  specifically for a significant increase  in the measured
blackbody radius in the burst rise and a following decrease, in bursts
where the X-ray flux remains almost constant at a peak value. Galloway
et  al.\ (2008a)  devised  a set  of  criteria based  on the  spectral
parameter variation in each burst  in order to identify PRE events and
to differentiate  them from other  typical X-ray bursts. We  adopt and
augment these criteria, as we discuss below.

\begin{figure}
\centering
   \includegraphics[scale=0.3, angle=0]{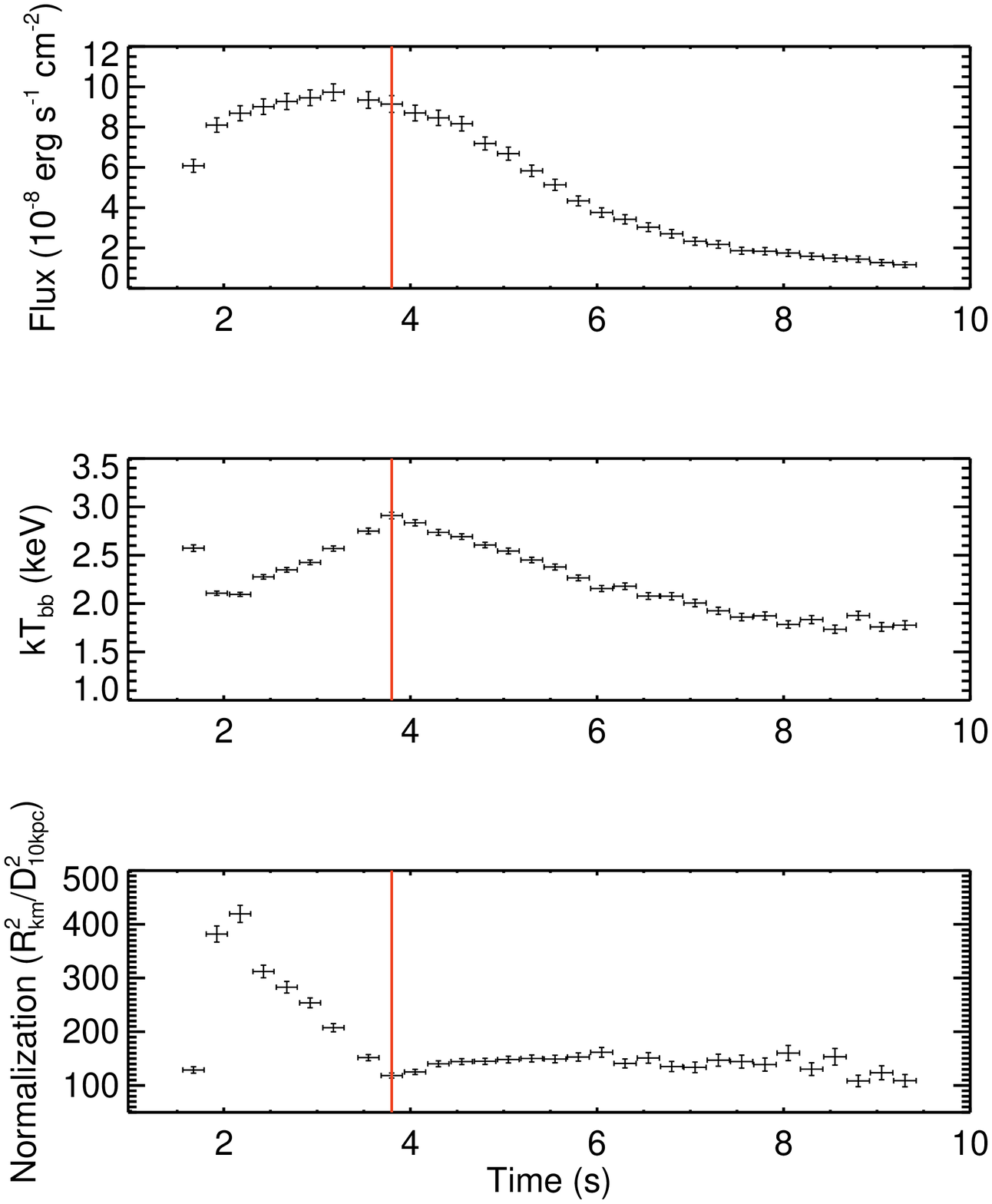}
   \includegraphics[scale=0.3, angle=0]{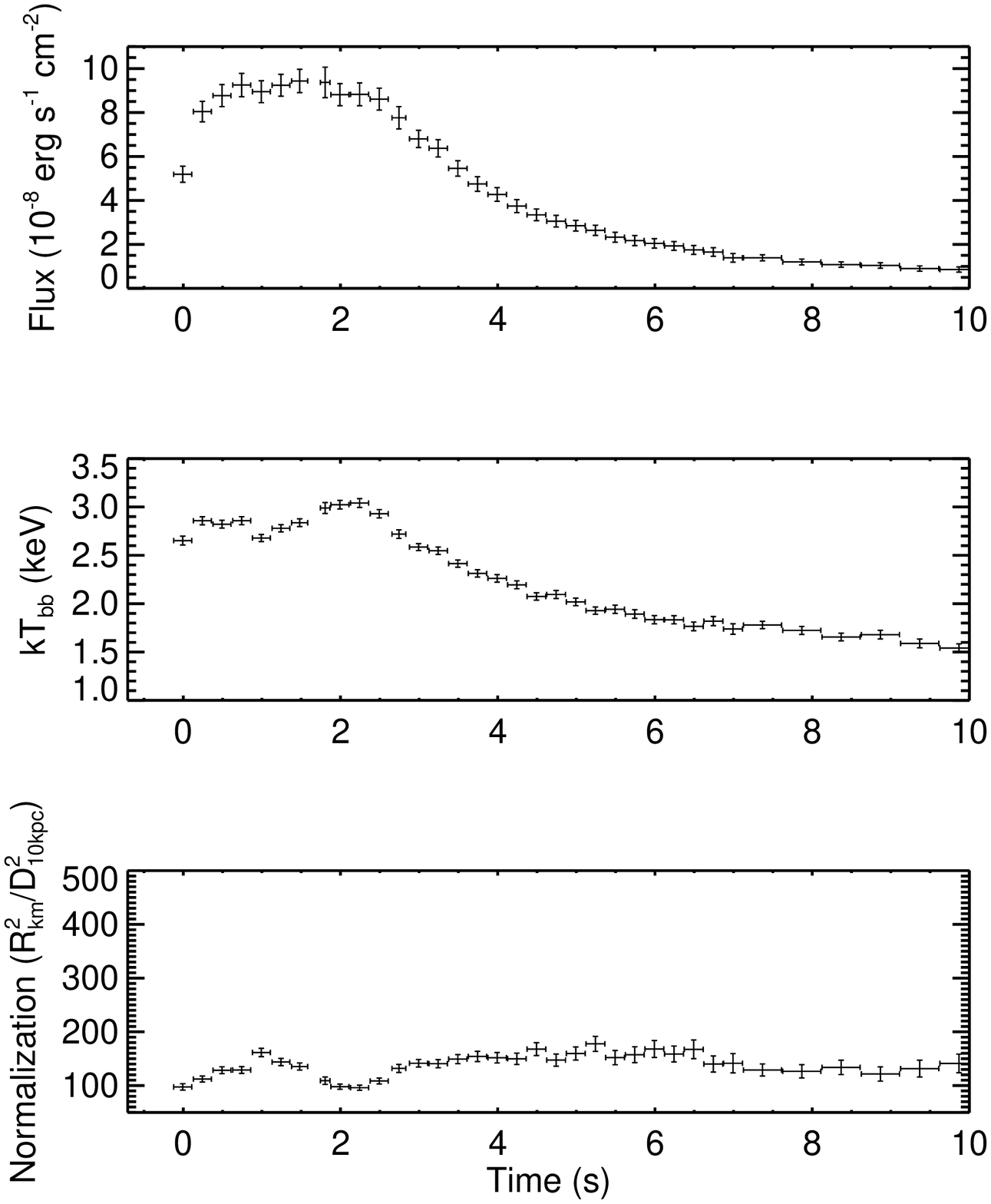}
   \caption{Examples of  X-ray bursts observed  from 4U~1728$-$34. The
     left panel shows burst \#86, which satisfies our criteria for PRE
     identification  summarized in  Section 3.  The right  panel shows
     burst \#104,  which does  not satisfy the  criteria hence  is not
     labeled as  a PRE event.   The selected touchdown moment  for the
     PRE event is also shown by a vertical line.}
\label{examples1}
\end{figure}

Galloway et al.\  (2008a) took the following measures  as the evidence
that a radius expansion  occurred: (1) the blackbody normalization $A$
reached a  (local) maximum close to  the time of peak  flux; (2) lower
values of  the normalization $A$ were measured  following the maximum,
with the decrease significant to 4 $\sigma$ or more; and (3) there was
evidence of a  (local) minimum in the fitted  temperature $T_c$ at the
same time as  the maximum in $A$.  In  Figure~\ref{examples1}, we show
examples  of  the spectral  evolution  of  two  different bursts  that
satisfy these criteria.  While the burst  in the left panel is a clear
PRE event, with  the photosphere at the peak  flux reaching many times
the neutron  star radius in the  cooling tail, the event  on the right
shows a higher normalization late in the burst than it does during the
assumed  photosphere expansion. In  fact, the  blackbody normalization
during  the  early  local  maximum  is  smaller  than  the  asymptotic
normalization of  even non-PRE bursts during their  cooling tails. We,
therefore, conclude that the latter example is not a secure PRE event.

In order  to eliminate  such cases, we  added an  additional criterion
that is  based on the  comparison of the peak  blackbody normalization
reached during an  X-ray burst, $A_{\rm peak}$, to  the measurement of
the average normalization, $A_{\rm cool}$, found from the cooling tail
for each  source. For the  former quantity, $A_{\rm peak}$,  we select
the peak  normalization that occurs  when the measured flux  is higher
than half  of the peak  flux.  This flux  limit ensures that  the peak
normalization is  selected when  the photospheric radius  expansion is
expected to  occur. For the  latter quantity, $A_{\rm cool}$,  we used
the average value  found from the cooling tails of  all the bursts for
each  source as  reported in  Paper I.   Note that  for  Aql~X$-$1 and
4U~0513$-$401, large  systematic uncertainties present  in the cooling
tails  prevented a  reliable measurement  of their  apparent  radii in
Paper   I.   Because   of  that,   we  used   approximate   values  of
$R/D=14.6$~km/10~kpc  and   $R/D=5.7$~km/10~kpc,  respectively,  which
correspond to the highest flux bins of their cooling tails.

\begin{figure}
\centering
   \includegraphics[scale=0.45, angle=270]{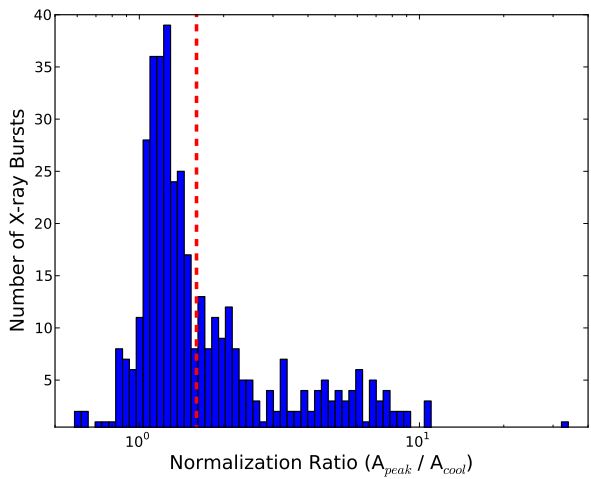}
   \caption{The  ratio  of  the  peak blackbody  normalization  values
     (A$_{\rm peak}$) found from all the X-ray bursts analyzed here to
     those obtained from the cooling tails (A$_{\rm cool}$) of all the
     X-ray  bursts  (Paper  I).   Larger  ratios  correspond  to  more
     distinguishable  photospheric  radius  expansion  episodes.   The
     dashed line shows our limit between the secure and non secure PRE
     events.}
\label{norm_histo}
\end{figure}

In  Figure~\ref{norm_histo},   we  show  the  histogram   of  all  the
normalization ratios  $A_{\rm peak}/A_{\rm  cool}$ for all  the bursts
observed from all  the sources included in this  study.  The resulting
histogram  shows  that the  distribution  of  the  ratio of  the  peak
normalization to the apparent radius  has a main peak around unity and
an extended tail towards higher  values. The high peak around unity at
the  peak normalization  shows that,  for  the majority  of the  X-ray
bursts, the  burning covers the  apparent surface area of  the neutron
star found  from the  cooling tails.  However,  there are a  number of
X-ray bursts where  the radius of the photosphere  reached values well
beyond  the apparent  neutron star  radius. We  consider these  as the
secure events where the photospheric radius expansion occurred.  Based
on this histogram, we tagged an  X-ray burst as a PRE event if $A_{\rm
  peak}/A_{\rm cool} > 1.65$. This value corresponds to the end of the
tail of the main peak in the histogram.

We excluded from  the final selected sample one  X-ray burst (burst \#
92) observed  from the direction  of 4U~1636$-$536.  Even  though this
burst satisfied the selection  criteria, the measured peak flux, $1.75
\times 10^{-8}$~erg~s$^{-1}$~cm$^{-2}$, is  much lower than the fluxes
reached in the rest of the burst sample and only half of the peak flux
reached in burst  ID 16, which is thought to be  a hydrogen rich burst
Galloway et al.\ (2006).   In Table~\ref{sourcestable}, we present the
number of PRE bursts for each  source that are obtained as a result of
the full set of criteria listed above. The additional criterion, which
eliminated  bursts  such  as the  one  shown  in  the right  panel  of
Figure~\ref{examples1}, naturally led to  numbers of secure PRE events
per source that  is somewhat lower than those  selected by Galloway et
al.\ (2008a). In addition, some of the difference in the number of PRE
events  is  caused by  the  $\gamma$ limit  we  imposed  in the  burst
selection  in   order  to   minimize  uncertainties  related   to  the
subtraction  of the  persistent  flux, which  we  take as  background.
Table~\ref{sourcestable} shows  the number  of bursts for  each source
that remain after the application of these criteria.

The number of  PRE events was most significantly  affected by the more
strict selection  criteria for 4U~1728$-$34:  16 out of the  69 events
that were tagged potentially as PRE by Galloway et al.\ (2008a) passed
the additional criteria.  This was  either because the increase in the
normalization was  not statistically significant when  compared to the
apparent radius of the neutron star  in the cooling tails of bursts or
because the normalization showed  a second increase during the cooling
tail  of the  burst  that sometimes  exceeded  the peak  normalization
during the  PRE phase, as in the  example shown in the  right panel of
Figure~\ref{examples1}.    X-ray  bursts   showing   similar  spectral
evolution were previously reported by  van Straaten et al.\ (2001) and
also by Galloway et al.\ (2003).  Given the fact that both at the peak
and during the cooling tails  of these bursts the normalization values
are  comparable to  the apparent  radius of  the neutron  star,  it is
possible that  the variation in the blackbody  normalization is caused
by a significant variation in the  color temperature and is not due to
a photospheric  radius expansion.  This  is also further  supported by
the fact that during the  peak of these particular X-ray bursts, color
temperatures were significantly higher than 2.5 keV and similar trends
in the  blackbody normalization  were also noted  in Paper I  at these
high  temperatures.  X-ray bursts  showing similar  spectral evolution
were also seen from 4U~1702$-$429 and Aql~X$-$1.

\begin{figure}
\centering
   \includegraphics[scale=0.35, angle=0]{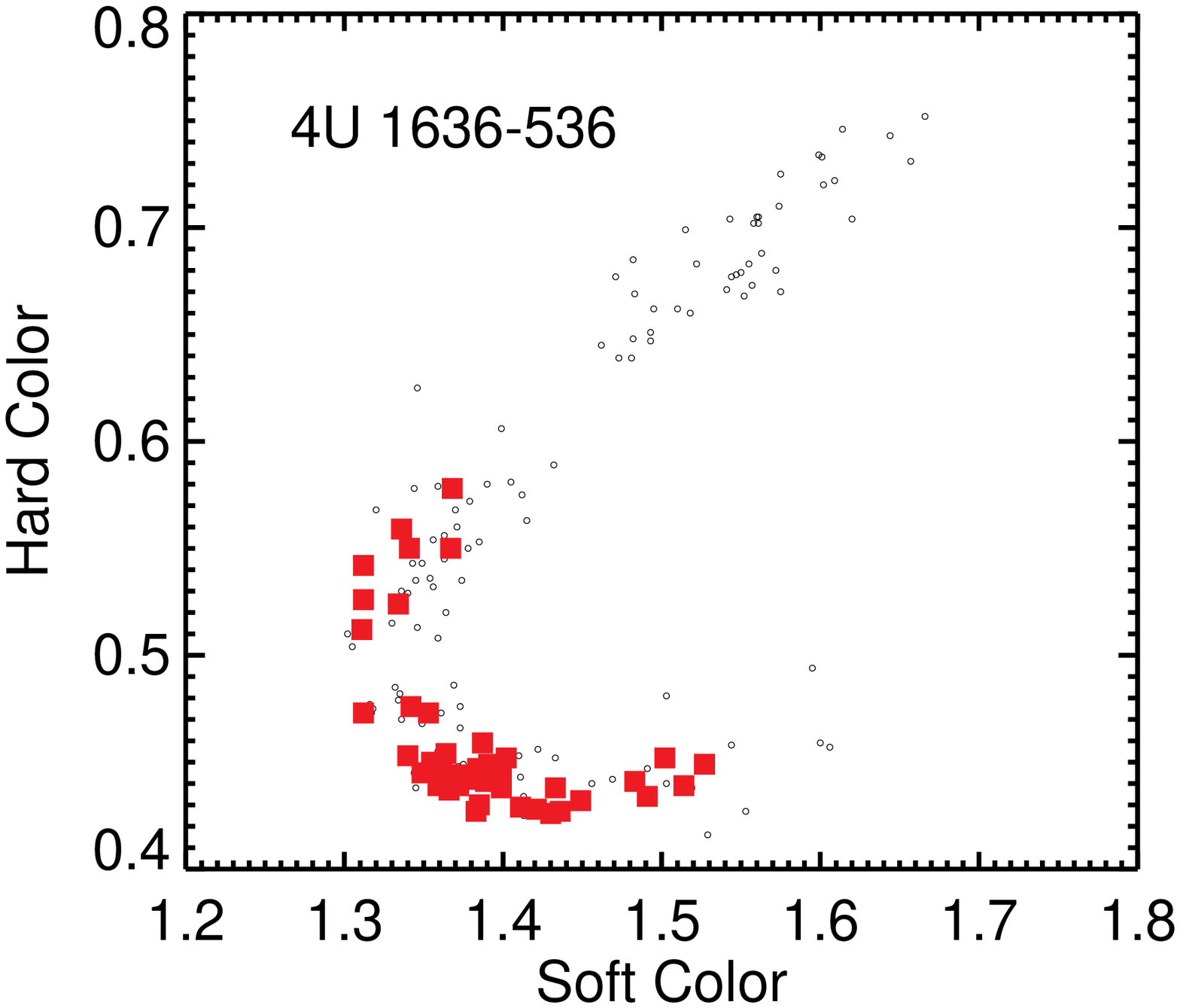}
   \includegraphics[scale=0.35, angle=0]{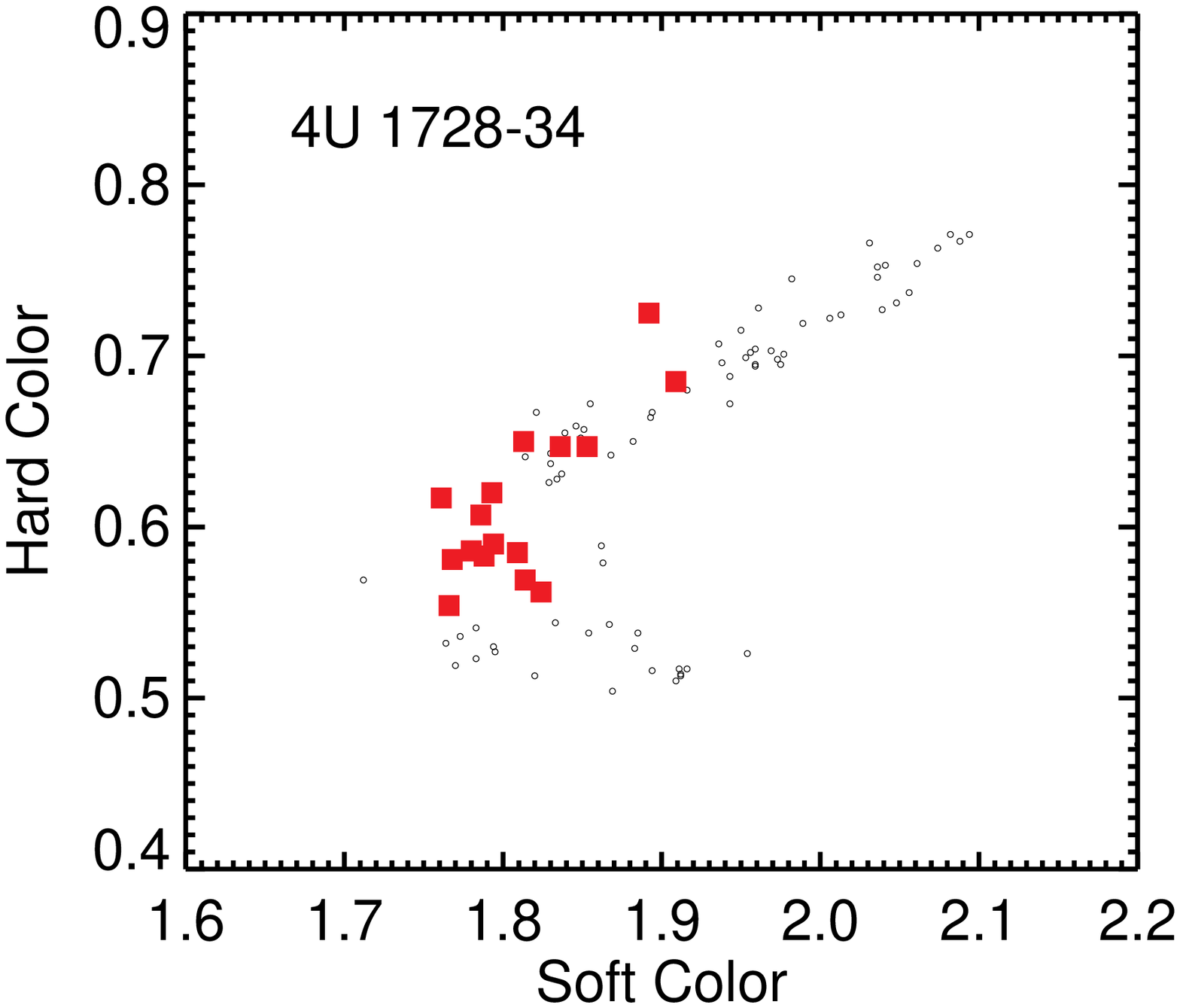} \caption{The
   positions of 4U~1636$-$536 (left panel) and 4U~1728$-$34 (right
   panel) on their color-color diagrams prior to the detection of an
   X-ray burst, using the data from Galloway et al.\ (2008a).  Red
   filled squares correspond to events that show clear evidence of
   photospheric radius expansion.  Secure PRE events appear to occur
   predominantly near the soft vertex of the color-color diagrams.}
\label{ccfig}
\end{figure}

We  finally explored  whether  PRE bursts  occur  only during  certain
spectral states of the neutron star binaries. To this end, we used the
data from Galloway et al.\ (2008a) to produce color-color diagrams for
the burst  sources and marked on  these diagrams the  locations of the
PRE and  non-PRE bursts.  Figure~\ref{ccfig}  shows the soft  and hard
color  for 4U~1728$-$34 and  4U~1636$-$536 prior  to the  detection of
each X-ray burst. The large (red) data points correspond to PRE bursts
while  the small (black)  points show  all other  thermonuclear bursts
from that  source. The PRE  bursts appear to occur  predominantly when
the sources lie  near the soft vertices of  their color-color diagrams
(see also  Muno et al.\ 2000).   However, the regions  with PRE bursts
still extend  across $\simeq  1/2$ of the  lengths of  the color-color
tracks. This minimizes the possibility that the reproducibility of the
inferred touchdown fluxes simply  reflects the fact we are considering
only very  similar X-ray  bursts in a  very narrow range  of accretion
rates.

Our  limit   on  the  pre-burst  flux,  i.e.,   the  requirement  that
$\gamma<0.1$,  excludes  the  brightest  regions  of  the  color-color
diagram of each source and may  also introduce a bias in our selection
of only particular PRE bursts.  This is not the case here, however, as
only a very  small fraction of the color-color  diagram of each source
corresponds  to $\gamma>0.1$  (compare, for  example,  the color-color
diagram  in Figure~\ref{ccfig}  to the  entire color-color  diagram of
4U~1728$-$34 in Figure~1 of Muno et al. 2002).

\section{Determination of the Touchdown Moment and the Eddington 
Limit}

We  now discuss  the determination  of  the touchdown  moment and  the
measurement of  the touchdown flux for  the PRE bursts  in our sample.
We  present here  the details  of the  analysis for  4U~1636$-$536 and
4U~1728$-$34, which  are the  sources with the  highest number  of PRE
events.

\begin{figure}
\centering
   \includegraphics[scale=0.4, angle=0]{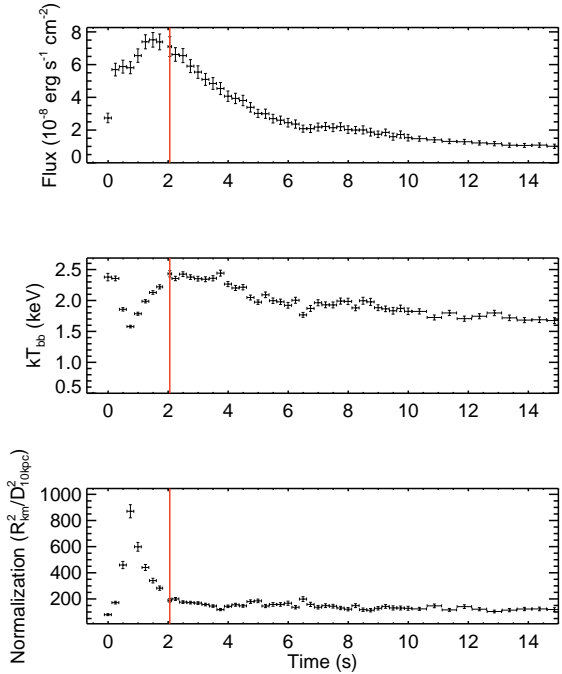}
   \caption{ An example X-ray burst observed from 4U~1636$-$536 (burst
     ID\# 150) where the touchdown moment is not defined at the moment
     when  the   temperature  reached  its  global   maximum  and  the
     normalization its  minimum but defined  as the first  moment when
     the temperature is within 1$-\sigma$ of the highest value.}
\label{examples2}
\end{figure}

The touchdown  moment is  defined as the  moment when  the photosphere
falls back onto  the neutron star, which is thought  to occur when the
observed   blackbody  normalization   reaches  its   lowest   and  the
temperature its highest value. In a very small number of X-ray bursts,
however, a  statistically insignificant temperature  maximum can occur
several seconds past the peak flux, as in the example of the PRE burst
from 4U~1636$-$536 shown in Figure~\ref{examples2}. In these cases, we
selected  the first  temperature maximum  (and  normalization minimum)
past the  peak flux,  ensuring that the  temperature at this  point is
within 1$-\sigma$  of its global  maximum. The touchdown moments  in a
total of 6 out  of 83 bursts from all of the  sources were selected in
this way.

The precise determination of the touchdown moment can also be affected
by data gaps  that are present in the science event  mode data in some
burst observations.  In these cases, where a gap may have an effect on
the  determination  of the  touchdown  moment,  we  checked whether  a
``burst   catcher''  mode  with   spectral  information   (e.g.,  mode
CB\_8ms\_64M\_0\_249\_H)  was used.  We found  that only  in  6 cases,
there were no  burst catcher mode data with  spectral information. For
the rest  of the X-ray bursts,  we made use  of the data in  the burst
catcher mode to determine the exact touchdown moments.

\begin{figure}
\centering         \includegraphics[scale=0.3,        angle=0]{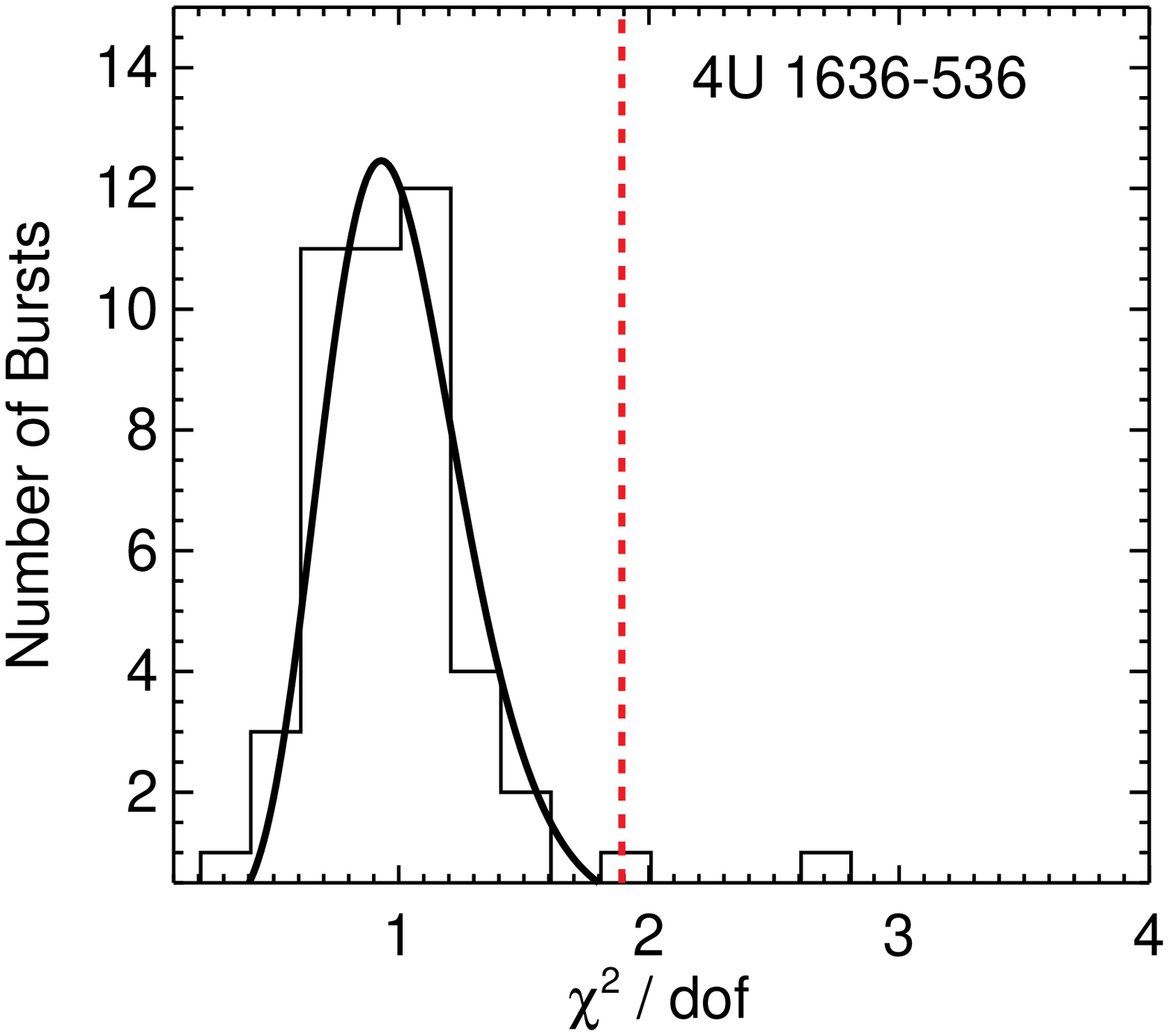}
\includegraphics[scale=0.3, angle=0]{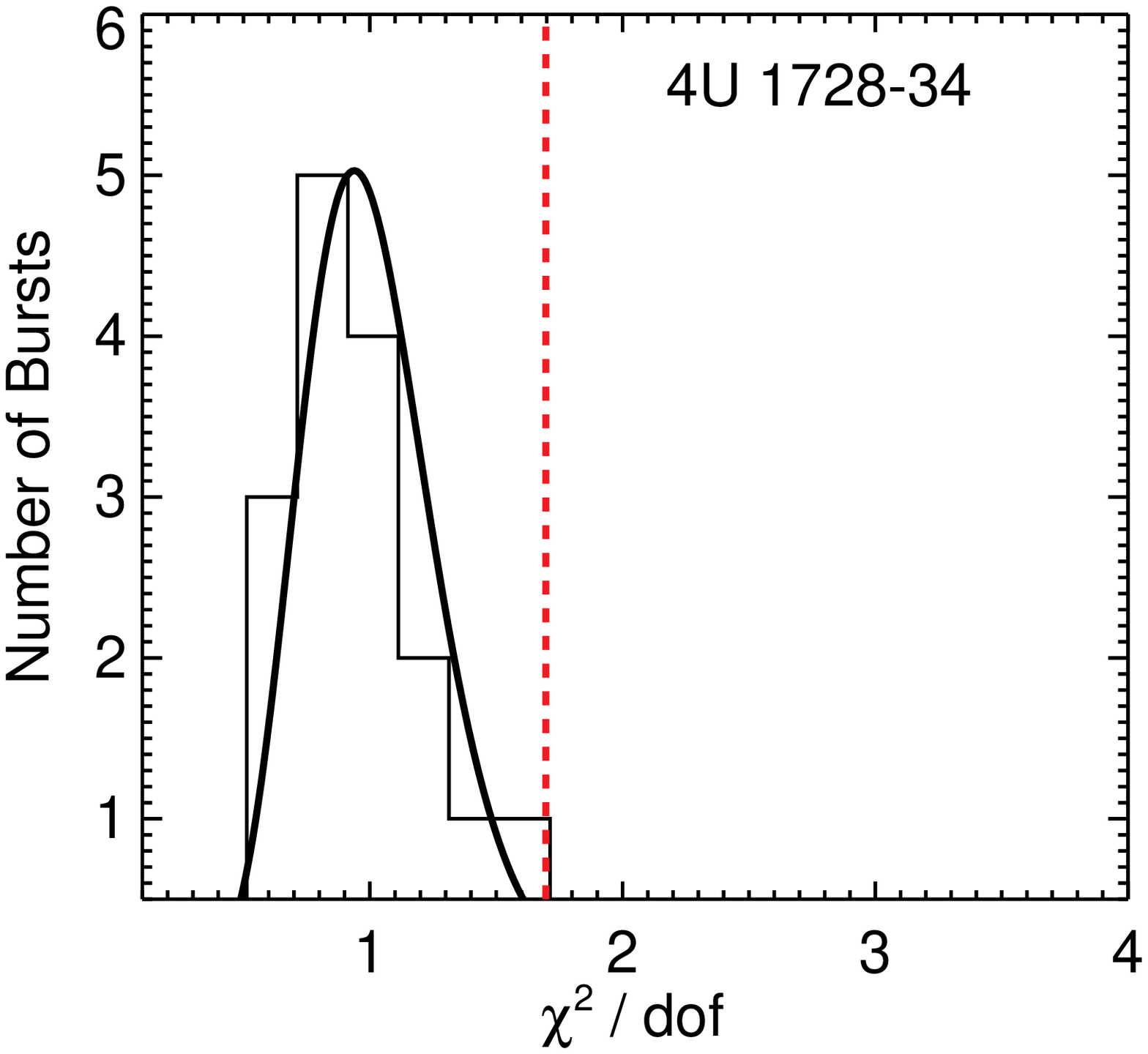} \caption{Distributions of
  $X^2$/dof  values obtained  from  model fits  to  the X-ray  spectra
  extracted  at the touchdown  moments of  X-ray bursts  observed from
  4U~1636$-$536  (left  panel) and  4U~1728$-$34  (right panel).   The
  solid  lines  show the  expected  distributions  for  the number  of
  degrees of freedom in the fits and the dashed lines show the highest
  values of $X^2$/dof that  was considered as statistically acceptable
  in Paper I  using the spectral fits of the cooling  tails of all the
  X-ray  bursts for  each  source. During  touchdown  the spectra  are
  described well by blackbody functions.}
\label{chi2}
\end{figure}

We fit the spectrum that we extracted at the touchdown moment for each
PRE  event  as  described  in  Section  2.   The  resulting  $X^2$/dof
histograms   for   4U~1636$-$536  and   4U~1728$-$34   are  shown   in
Figure~\ref{chi2}   (see   Paper~I   for   the  definition   of   this
statistic). Using  the $X^2$/dof limits determined in  Paper~I, we can
determine whether  a particular fit is statistically  acceptable or it
should be  excluded from further  analysis.  The X-ray spectra  at the
touchdown  moments  were  well  described  with  blackbody  functions,
leading in general to small $X^2$/dof values.  Therefore, applying the
$X^{2}$/dof  limits forced  us to  exclude only  one X-ray  burst from
4U~1705$-$44  (burst \#  1) and  two X-ray  bursts  from 4U~1636$-$536
(bursts \# 3 and 9).

\section{Systematic Uncertainties in the Eddington Limit}

In  this   section,  we  will   address  the  formal   and  systematic
uncertainties in the touchdown fluxes  obtained from the PRE bursts of
each source.  As  before, we will first focus on  the two sources with
the highest number of bursts to  present the details of the method and
then extend our analysis to the  rest of the sample.  We will start by
discussing our  determination of the bolometric flux  at touchdown and
its formal uncertainty. We will then explore whether the different PRE
bursts  from the  same  source  reach a  touchdown  flux that  remains
statistically constant between bursts.

For each burst, the bolometric  flux at touchdown is obtained from the
combination of the blackbody  temperature and normalization.  Figure 6
shows  the  68\%  and   95\%  confidence  contours  of  the  blackbody
normalization and temperature inferred  from fitting the X-ray spectra
obtained   during   the   touchdown   moment  for   4U~1728$-$34   and
4U~1636$-$536.   We also plot  in these  figures contours  of constant
bolometric  flux,  shown  as  dotted  (red) lines.   Even  though  the
uncertainties in the normalization and temperature are correlated, the
bolometric flux  in each burst  is well constrained.   Furthermore, as
Figure  6 shows, the  individual confidence  contours from  each burst
appear to  be in very good  statistical agreement with  each other for
both sources.

\begin{figure}
\centering
   \includegraphics[scale=0.4, angle=0]{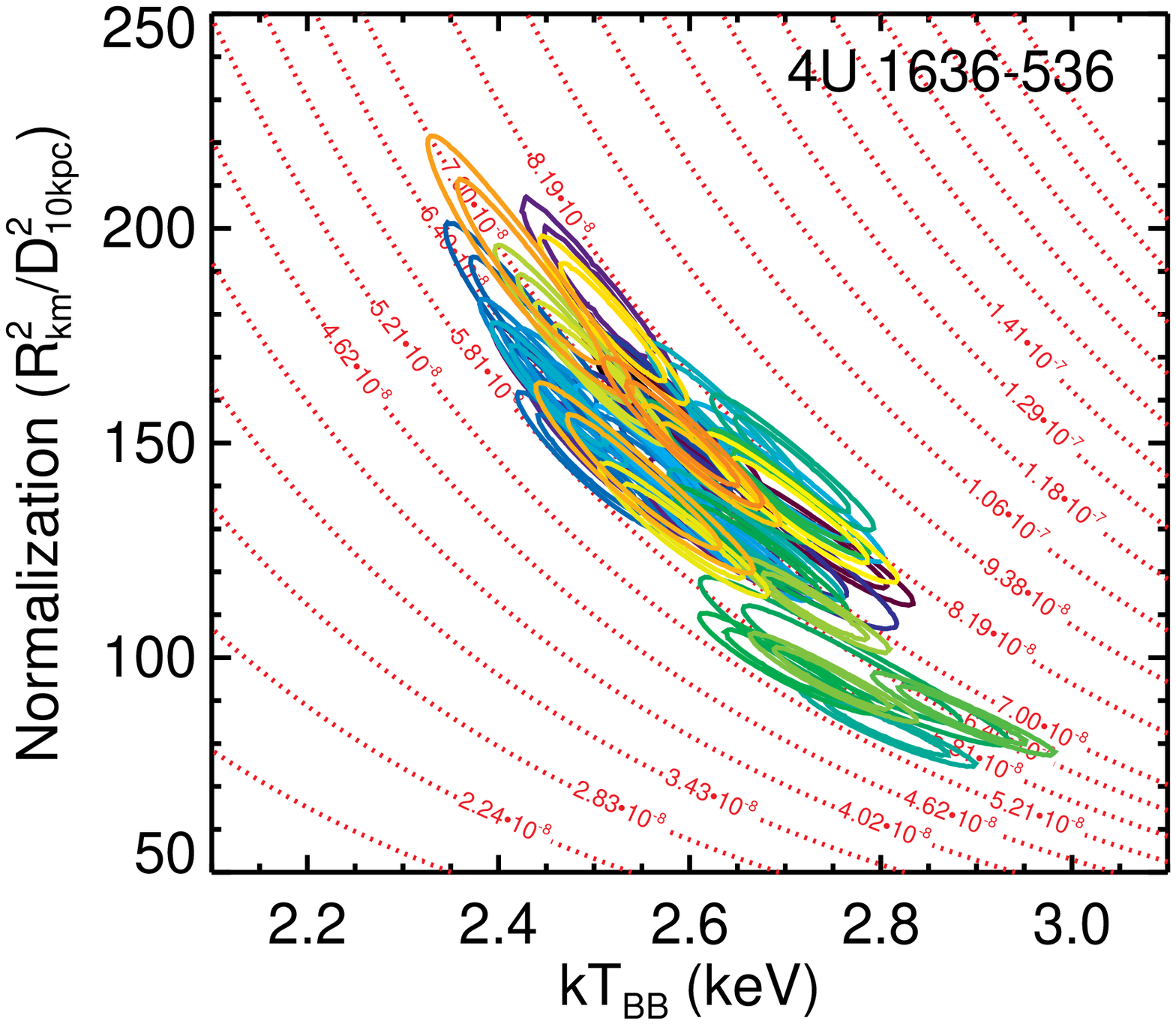}
   \includegraphics[scale=0.4, angle=0]{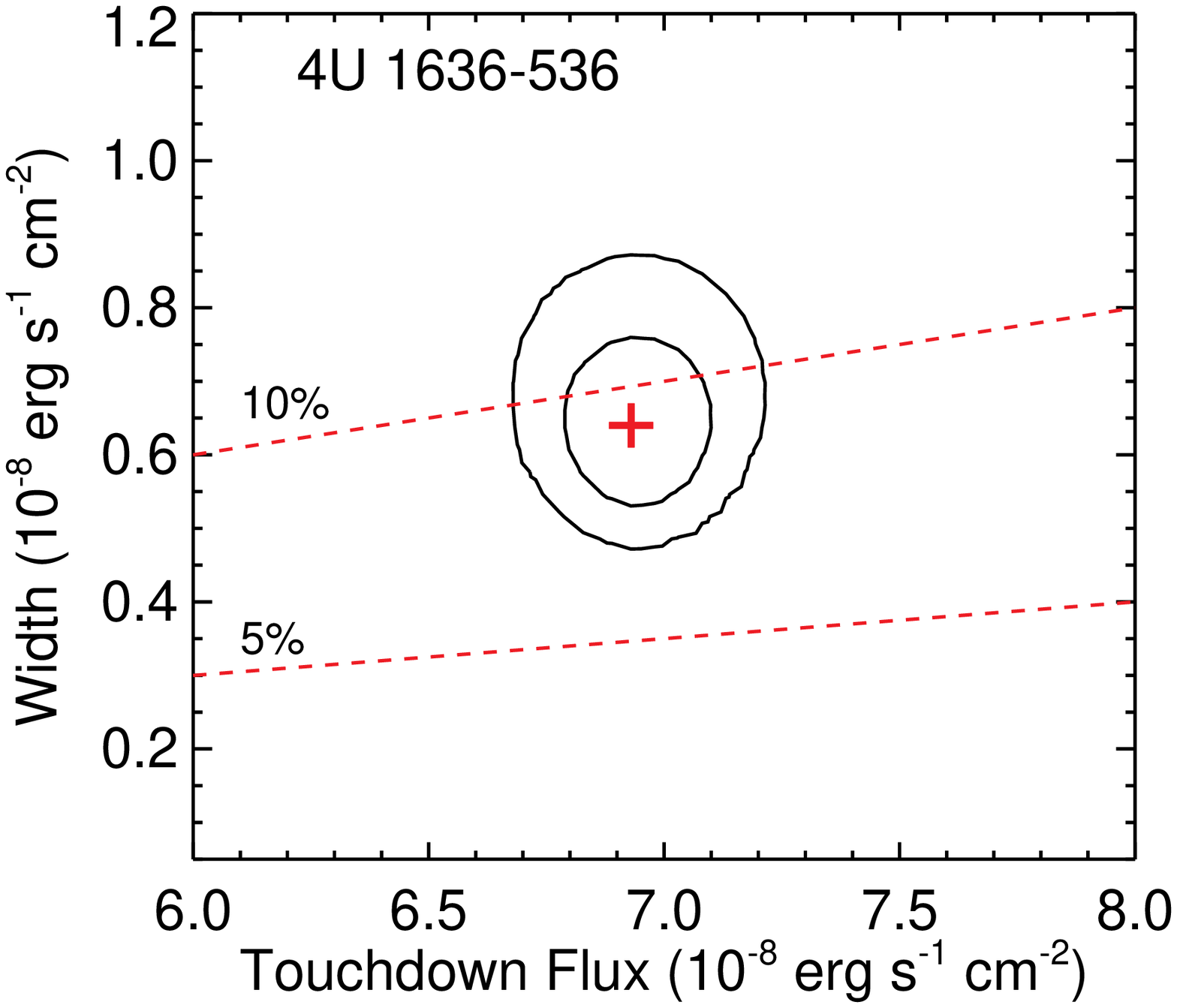}
   \includegraphics[scale=0.4, angle=0]{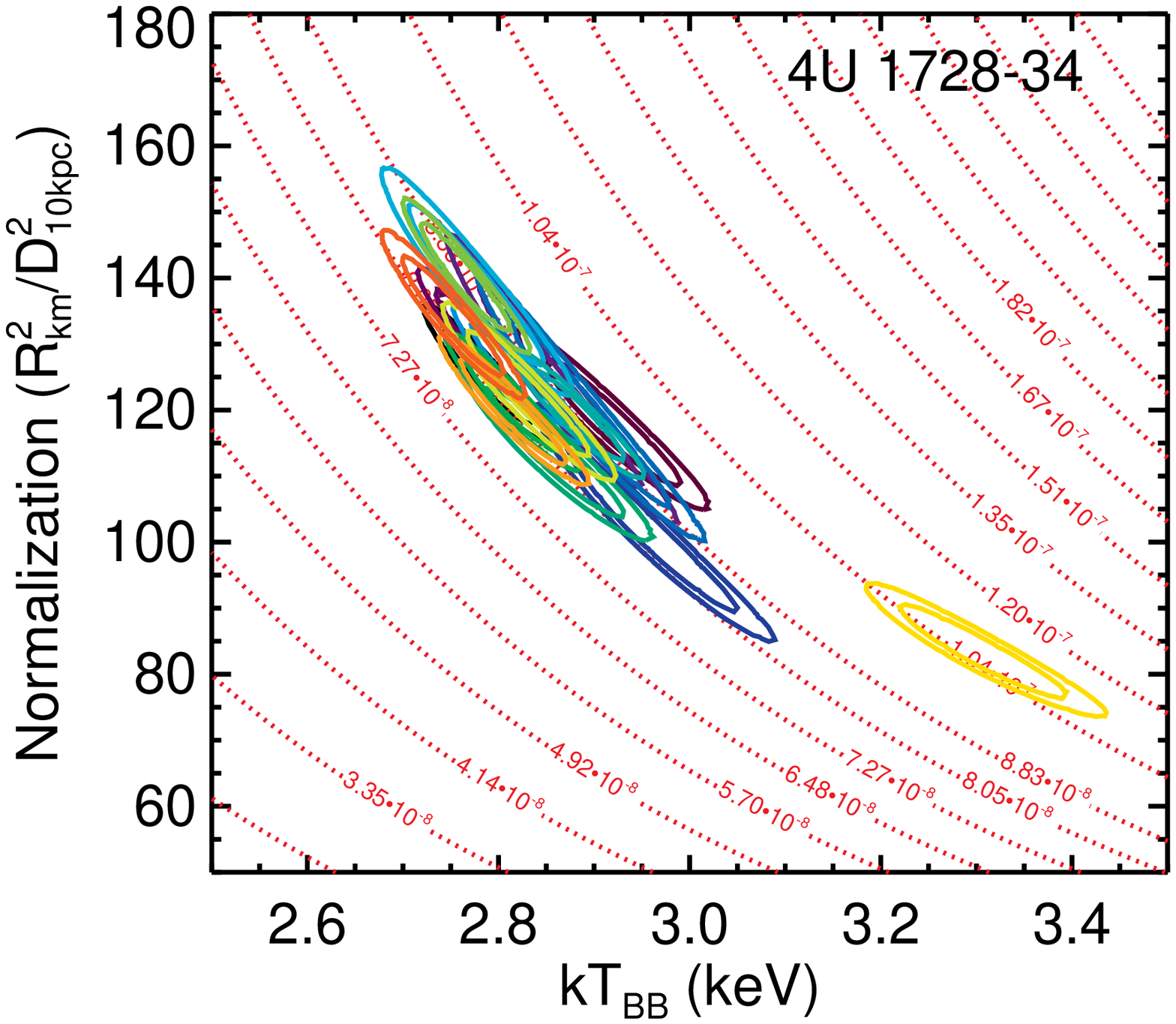}
   \includegraphics[scale=0.4, angle=0]{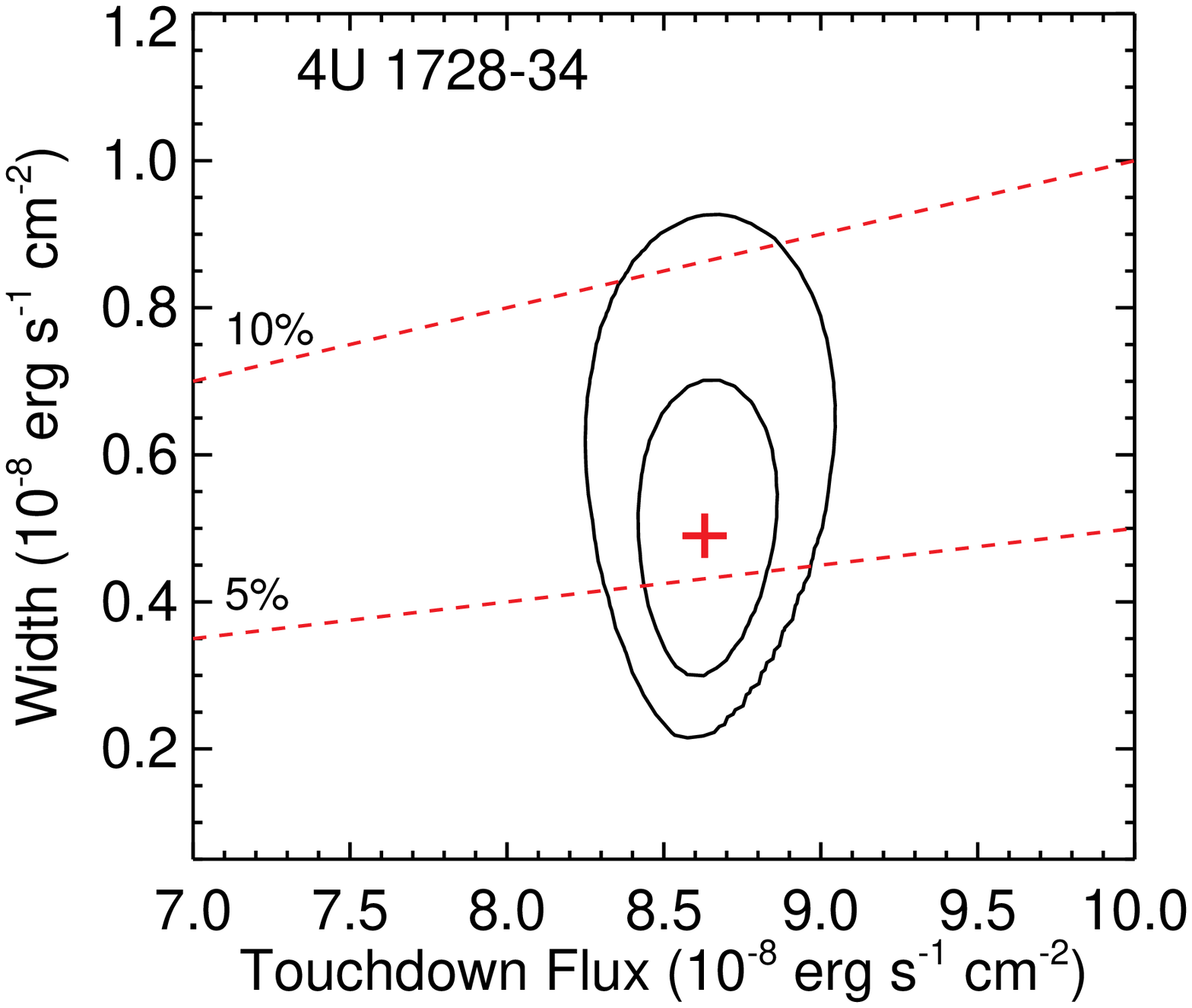} 
\caption{{\it Left Panels :} 68\% and 95\% confidence contours of the blackbody
   normalization and temperature obtained from fitting the X-ray
   spectra at the touchdown moments of each PRE burst observed from
   4U~1636$-$536 and 4U~1728$-$34.  The dotted red lines show contours
   of constant bolometric flux. {\it Right Panels:} 68\% and 95\%
   confidence contour of the parameter of an assumed underlying
   Gaussian distribution of touchdown fluxes.  The width of the
   underlying distribution reflects the systematic uncertainty in the
   measurements. The dashed red lines show the width when the
   systematic uncertainty is 5\% and \%10 of the mean touchdown flux.}
\label{tdex}
\end{figure}

The  distribution  of  inferred  bolometric  fluxes  at  touchdown  is
expected  to   have  a  finite  width  both   because  of  measurement
uncertainties and  because of the possible variations  in the physical
conditions that  determine the emerging  flux during a PRE  burst. The
measurement uncertainties  include formal uncertainties  from counting
statistics,  the  uncertainties  in  the  bolometric  correction,  the
subtraction of  the background emission, and the  determination of the
touchdown  moment.   Anisotropies in  the  bursts,  variations in  the
composition and the reflection  off the accretion flow (e.g., Galloway
et al.\ 2004, 2006), and variations in the Compton upscattering in the
converging  inflow  prior  to  touchdown  are  some  of  the  physical
mechanisms that can contribute to the intrinsic spread.

For the high temperatures observed  during the touchdown phases of the
bursts, most of the burst spectrum falls within the RXTE energy range,
resulting in bolometric corrections that  are at most 7\% (Galloway et
al.\ 2008a). Therefore, any uncertainties in the bolometric correction
can  only  introduce minimal  spread  to  the  width of  the  observed
touchdown fluxes. Uncertainties in  the determination of the touchdown
moment are also expected to be  of the same magnitude since the fluxes
in the nearby time bins differ typically by less than 10\% (see, e.g.,
Figures 2  and 4).   Our 10\% limit  on the pre-burst  persistent flux
bounds  the  uncertainties  introduced   by  our  subtraction  of  the
background.  We can  also estimate the expected variations  due to the
Compton  upscattering in the  converging flow:  this effect  scales as
$v/c$  and can,  therefore, introduce  an uncertainty  at most  of the
order of  10\% (van Paradijs and  Stollman\ 1984). On  the other hand,
variations in  the isotropy or the  composition of the  bursts can, in
principle, generate larger spread in the touchdown fluxes.

Our goal is to quantify  the widths of the underlying distributions of
touchdown fluxes, which we will call systematic uncertainties that are
potentially caused by any of  these effects. In order to achieve this,
we need  an approach that is valid  both in the limit  when the formal
uncertainty for each measurement is  much smaller than the variance of
the distribution  of their central values  as well as  in the opposite
extreme.   In the  first  case, the  variance  of the  mean values  is
practically equal to the width  of the underlying distribution. In the
opposite limit, when the  formal uncertainties in each measurement are
comparable or larger  than the variance of the  mean values, one could
compute the  systematic uncertainty $\sigma_{\rm  sys}$ by subtracting
in  quadrature the  formal  uncertainty $\sigma_{\rm  form}$ from  the
variance $\sigma_{\rm var}$, i.e.,
\begin{equation}
\sigma_{\rm sys}^2 = \sigma_{\rm var}^2 - \sigma_{\rm form}^2. 
\end{equation}
This can be carried out only if the formal uncertainties in each 
measurement are the same.

In our sample,  however, each flux measurement has  a different formal
uncertainty  and  the uncertainty  in  each  measurement is  sometimes
comparable to and sometimes smaller than the variance of the mean flux
values for different  sources. In order to properly  account for this,
we follow here the Bayesian  analysis method discussed in Paper I that
allows us to determine the intrinsic spread of touchdown fluxes.

\begin{figure}
\centering
   \includegraphics[scale=0.4, angle=0]{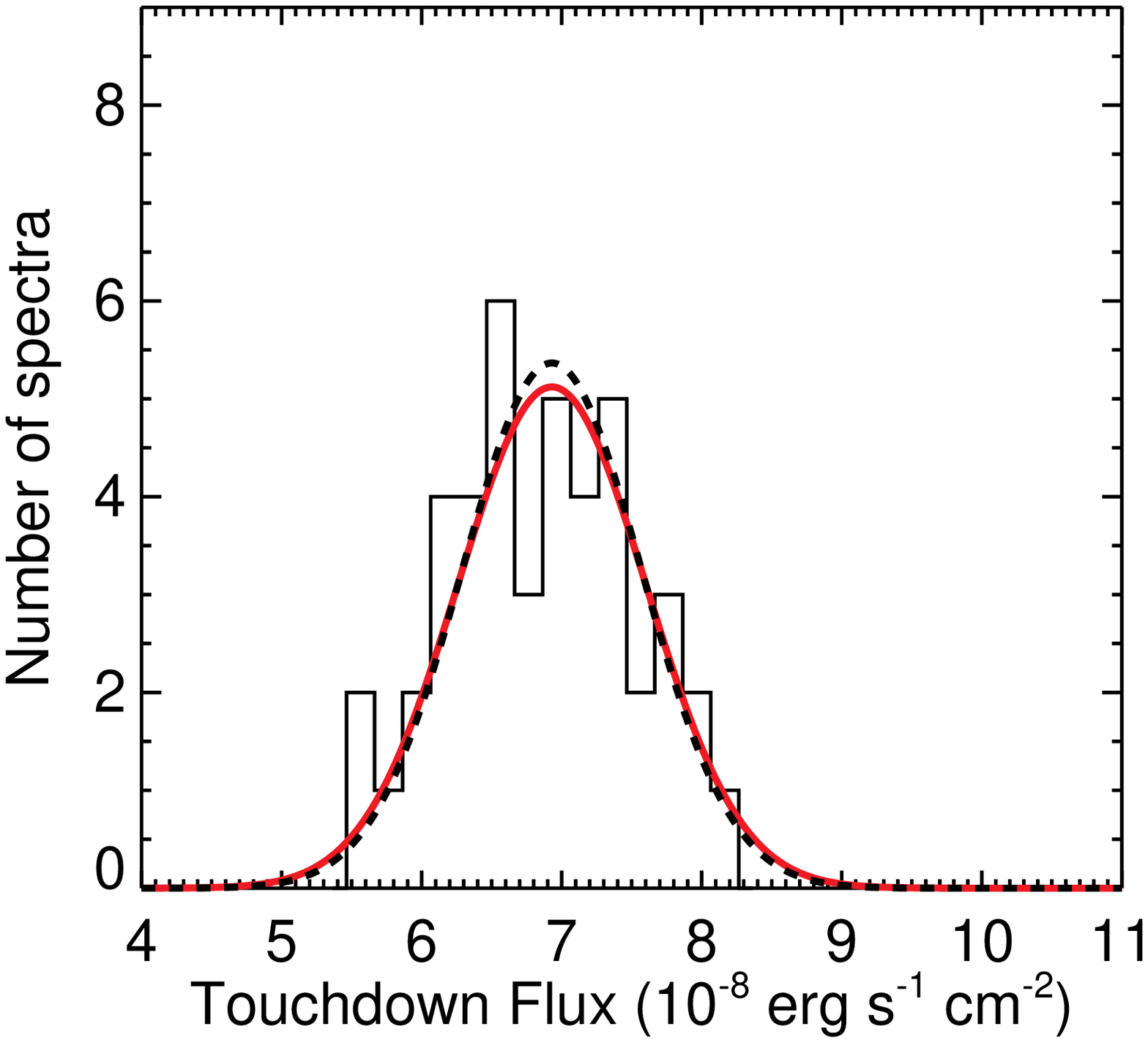}
   \includegraphics[scale=0.4, angle=0]{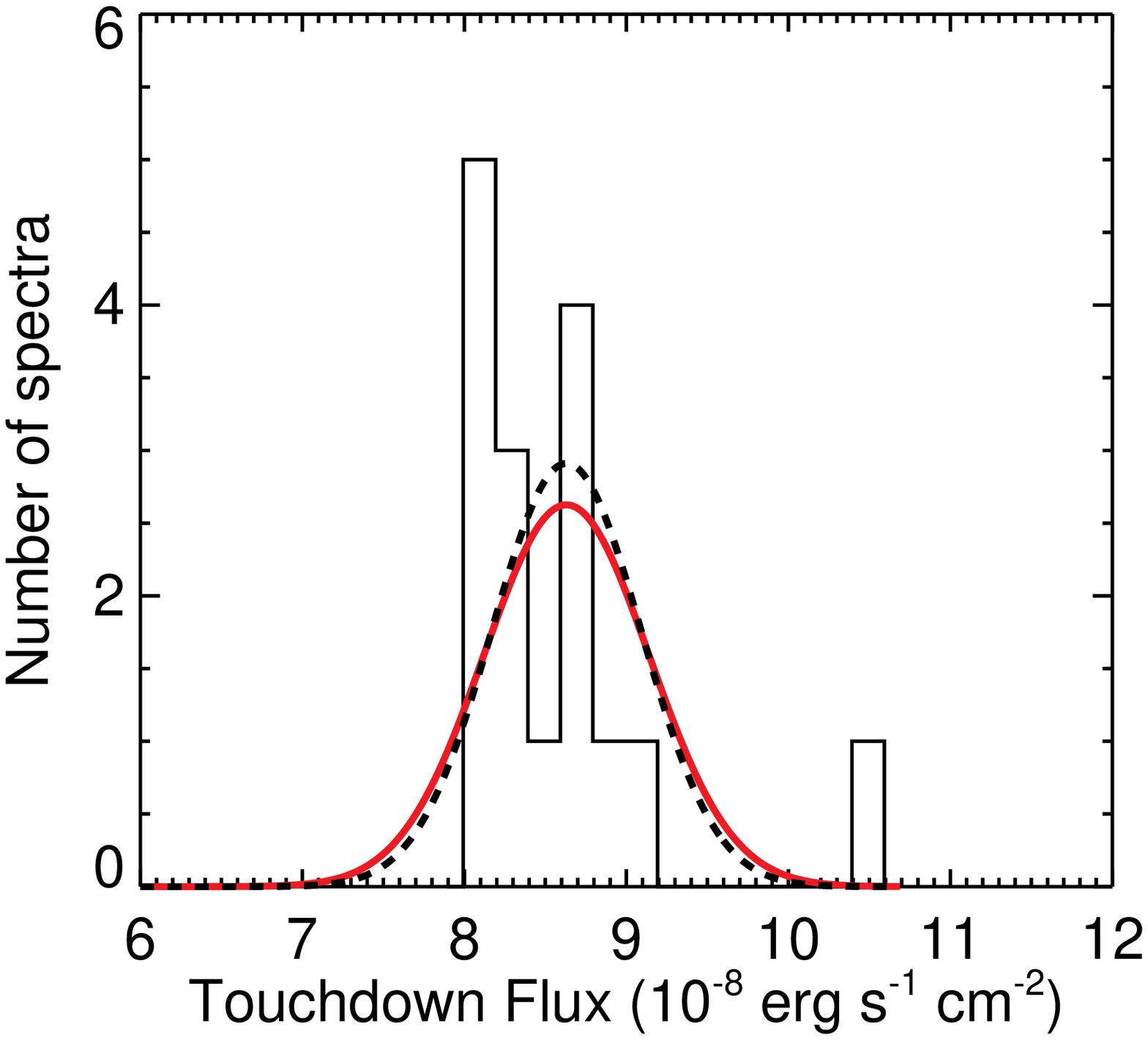} \caption{Histogram of
   measured touchdown fluxes for 4U~1636$-$536 (left panel) and
   4U~1728$-$34 (right panel).  The red solid line shows the
   underlying Gaussian distribution as inferred from the Bayesian
   analysis. The black dashed curve shows the distribution of the
   touchdown fluxes when the observational uncertainties are taken
   into account.  The width of the underlying distribution reflects
   the systematic uncertainty in the measurements.}
\label{tdsys}
\end{figure}

In the Bayesian analysis,  we first determine the formal uncertainties
of the measured bolometric fluxes for each burst and each source using
the confidence contours shown in Figure~\ref{tdex} and report these in
Table~\ref{tdres}. We  model the underlying  distribution of touchdown
fluxes as  a Gaussian. The  observed distribution is a  convolution of
the underlying  distribution with the  individual formal uncertainties
for  each burst  that we  measured above.   We then  use  the Bayesian
technique presented  in Paper I  to determine the most  probable value
$F_0$ and  width $\sigma$ of the underlying  distribution of touchdown
fluxes  for each  source.  Figure~\ref{tdsys}  shows the  histogram of
observed  touchdown fluxes  as well  as the  most  probable underlying
distribution for the two  sources 4U~1636$-$536 and 4U~1728$-$34.  The
right panels  of Figure~\ref{tdex}  show the full  confidence contours
for the parameters of these underlying distributions.  Even though the
most probable values for the touchdown fluxes can be determined within
a few percent, there is clear evidence for a 5\%-10\% spread, which we
attribute to the physical mechanisms discussed above.

In Figures~8$-$9 and Table~\ref{tdaverage}, we show the results of the
same  analysis for all  the other  sources.  Naturally,  the intrinsic
widths in  the touchdown fluxes for  the sources with  very few bursts
are more difficult to determine. In all cases except Aql~X-1, however,
the  level of systematic  uncertainties is  not inconsistent  with the
5-10\% level inferred  for the two sources with  many bursts. Notably,
Aql X-1 is also the source  with the largest variation in the apparent
surface area during the cooling tails of its bursts (Paper I).

\begin{figure}
\centering
   \includegraphics[scale=0.3, angle=0]{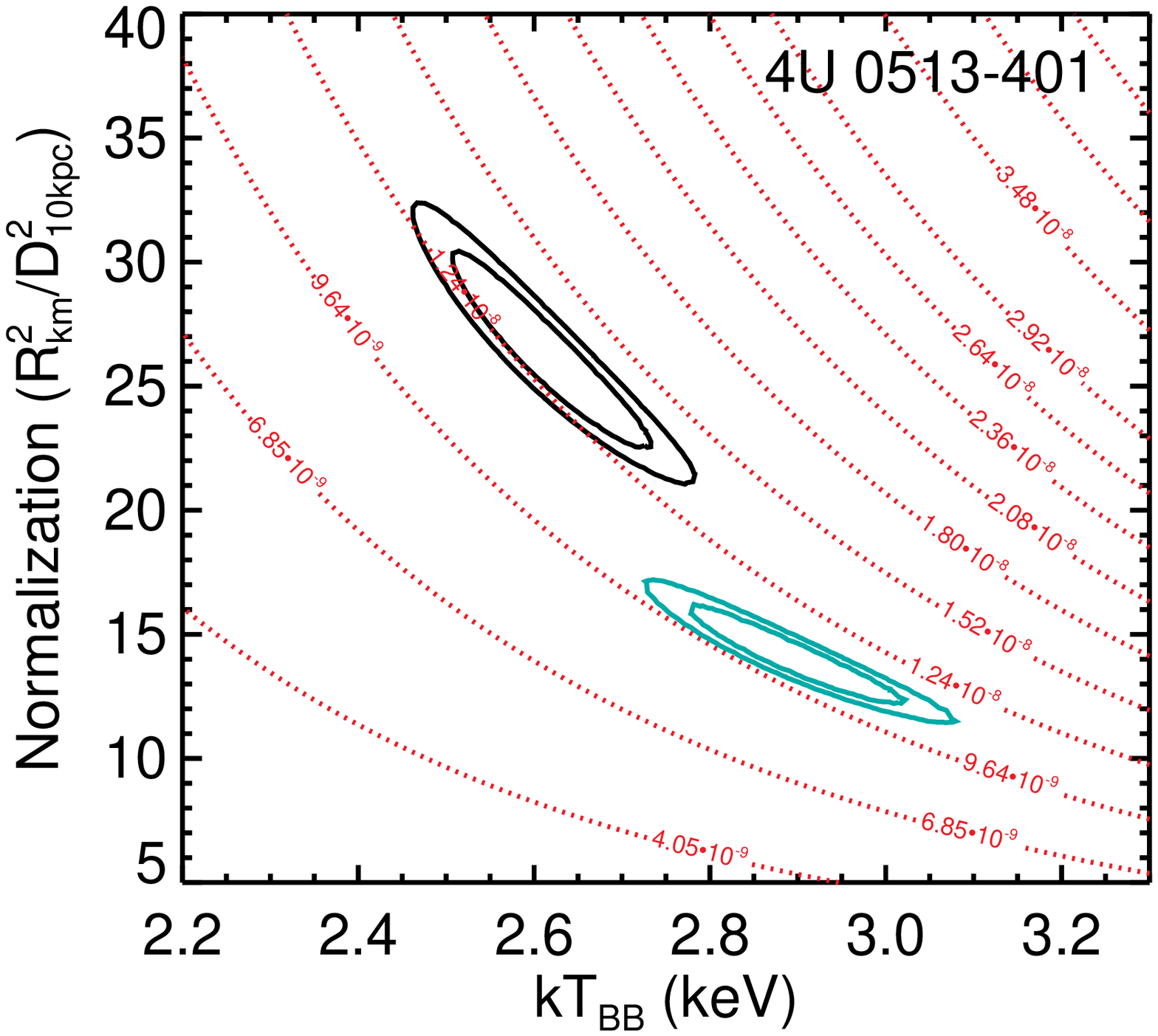}
   \includegraphics[scale=0.3, angle=0]{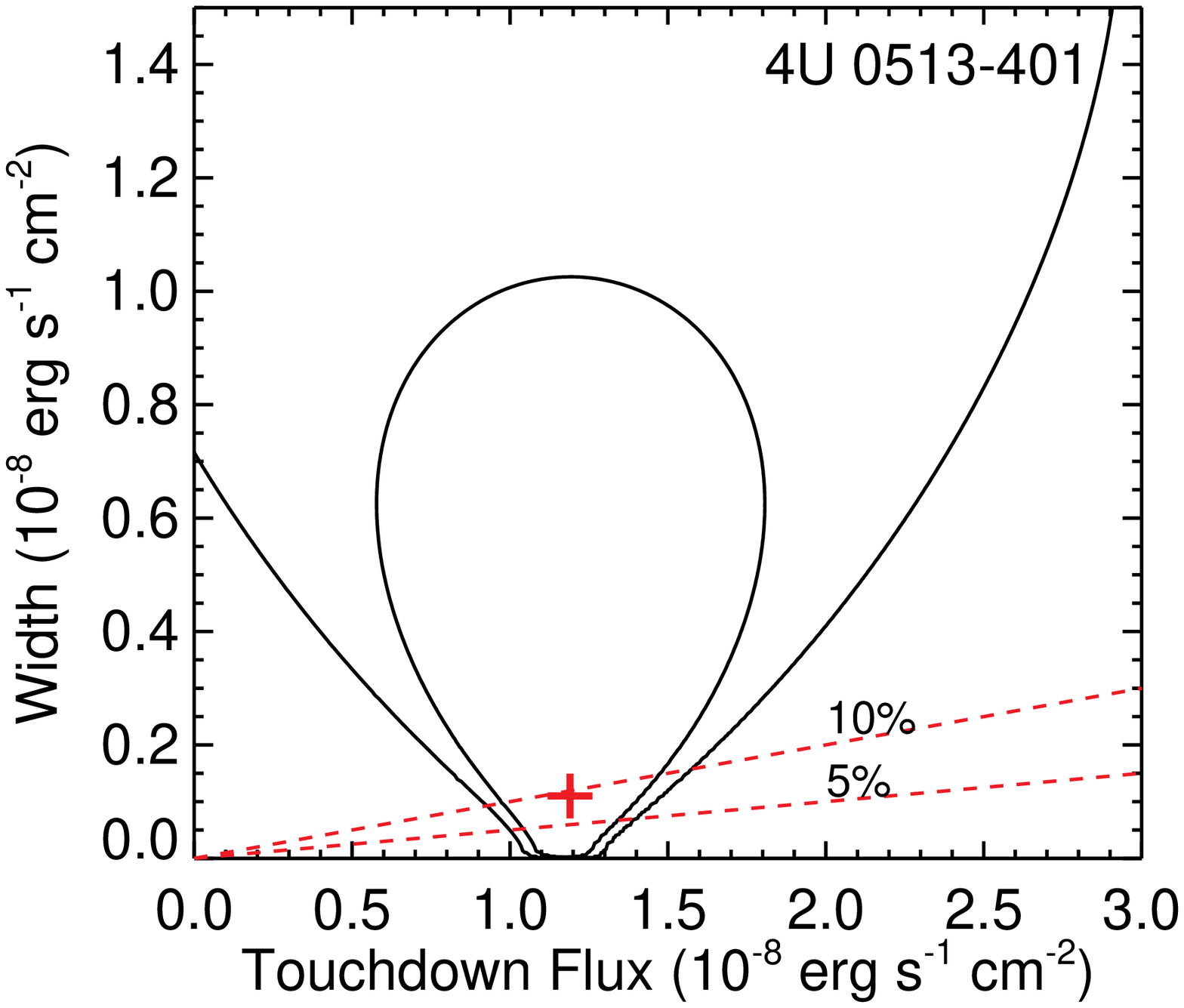}
   \includegraphics[scale=0.3, angle=0]{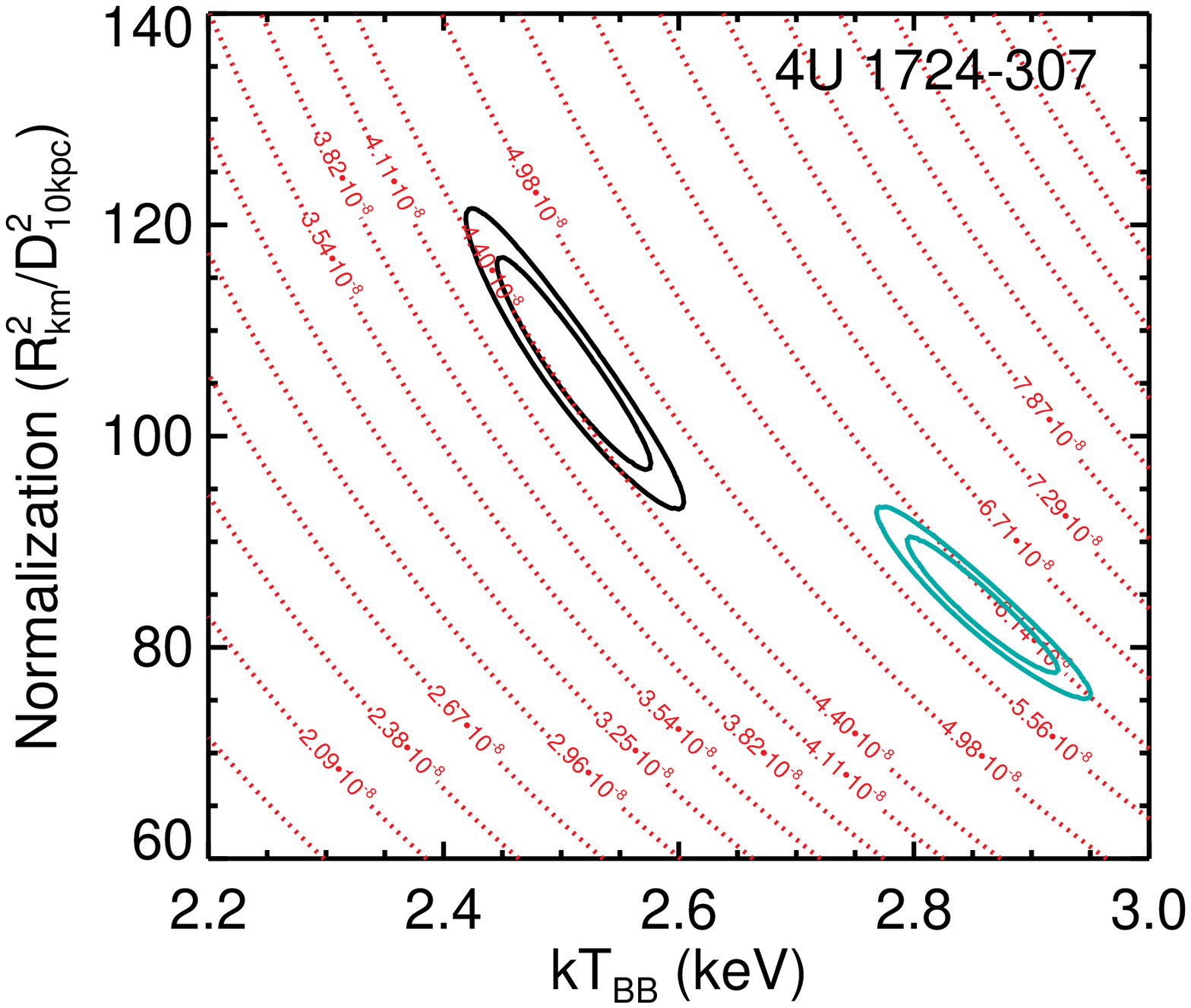}
   \includegraphics[scale=0.3, angle=0]{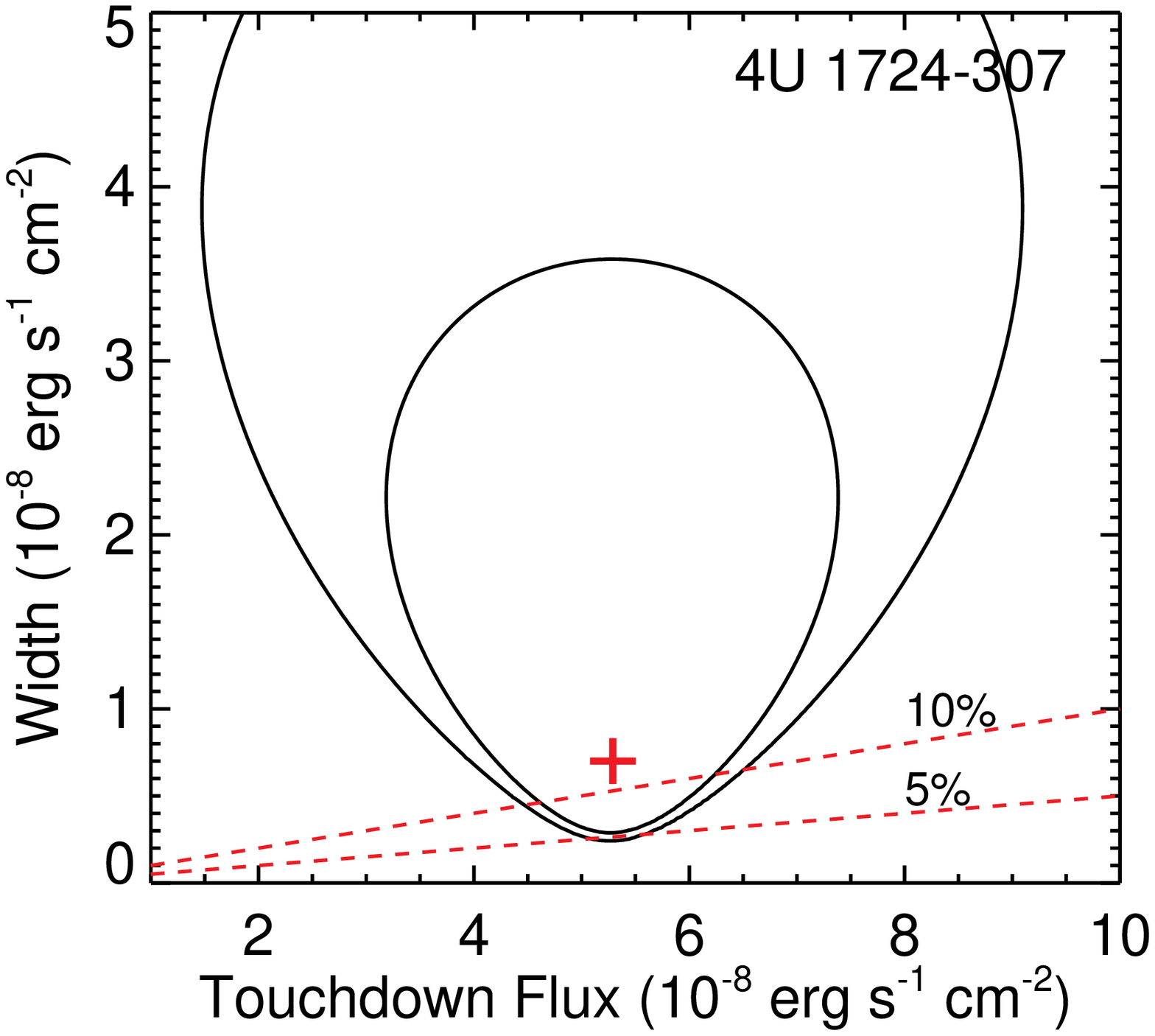}
   \includegraphics[scale=0.3, angle=0]{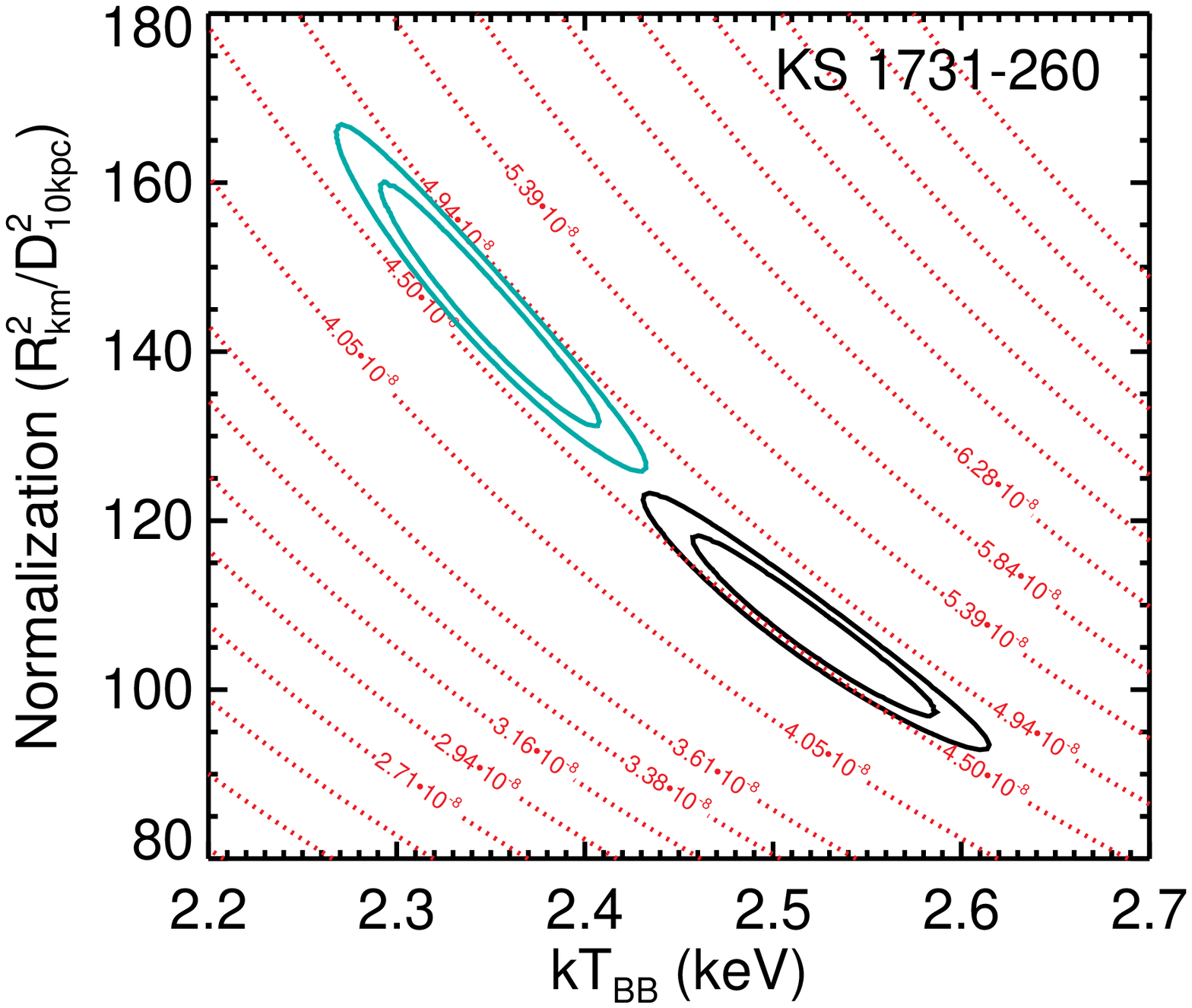}
   \includegraphics[scale=0.3, angle=0]{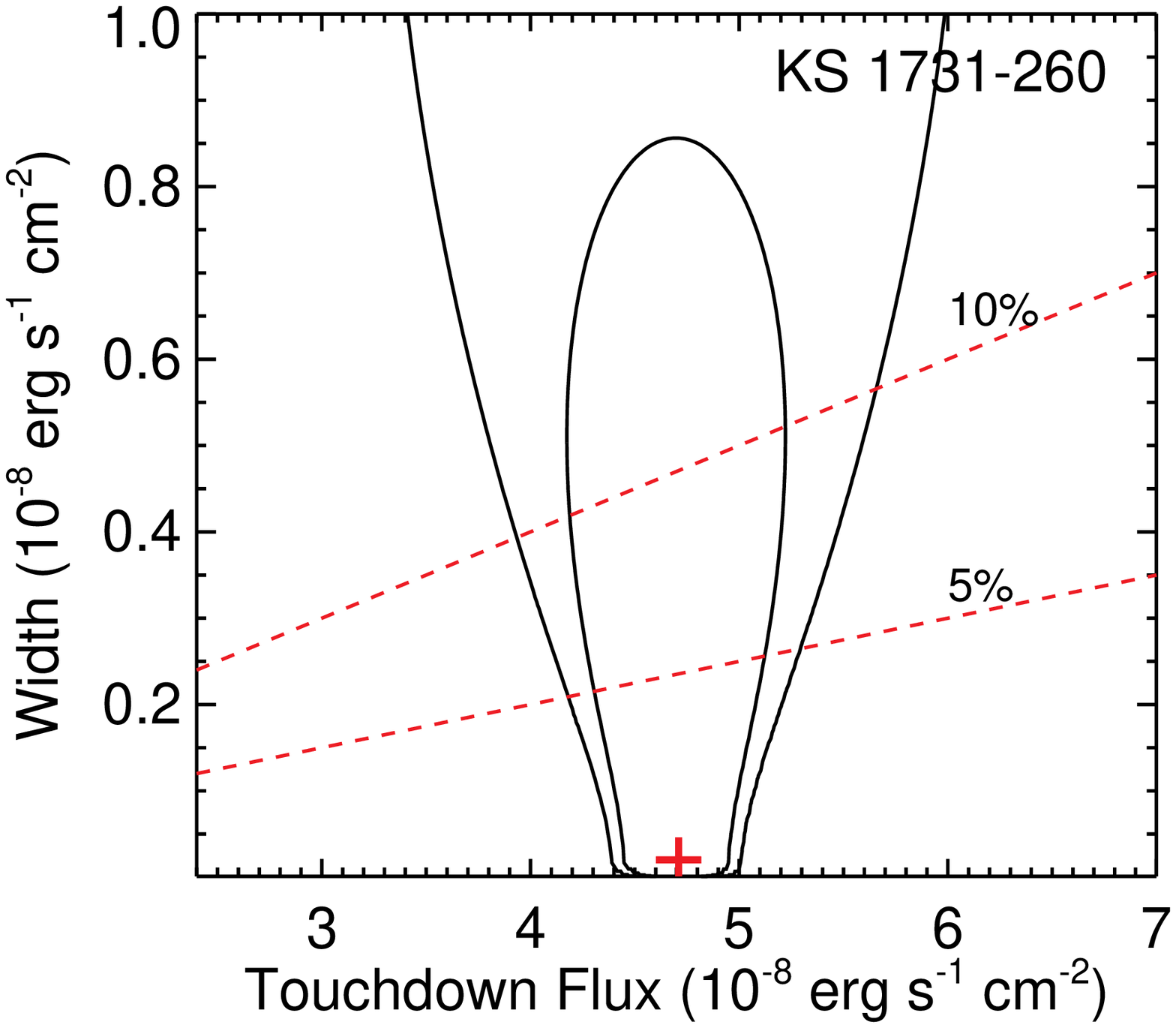}
   \includegraphics[scale=0.3, angle=0]{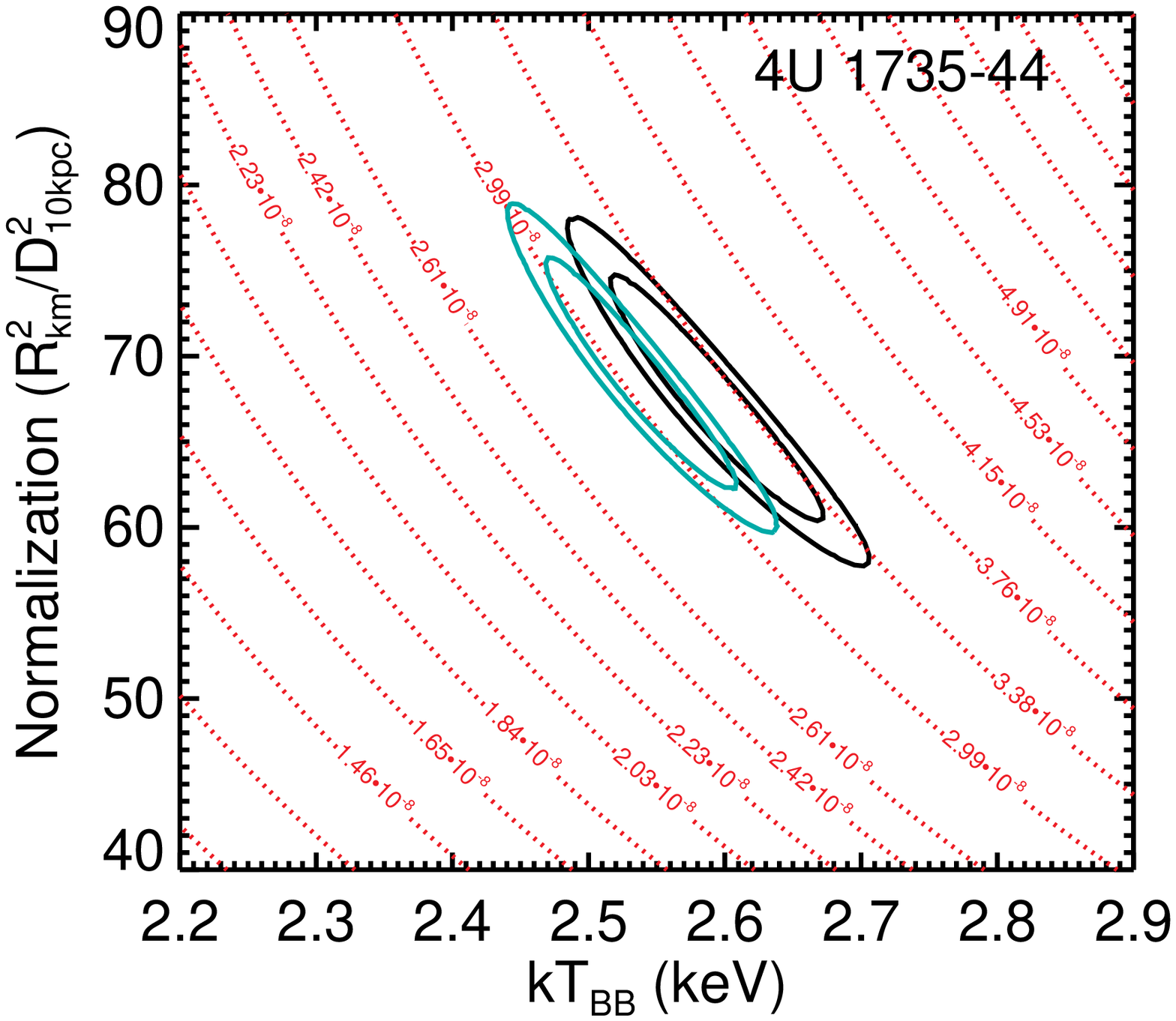}
   \includegraphics[scale=0.3, angle=0]{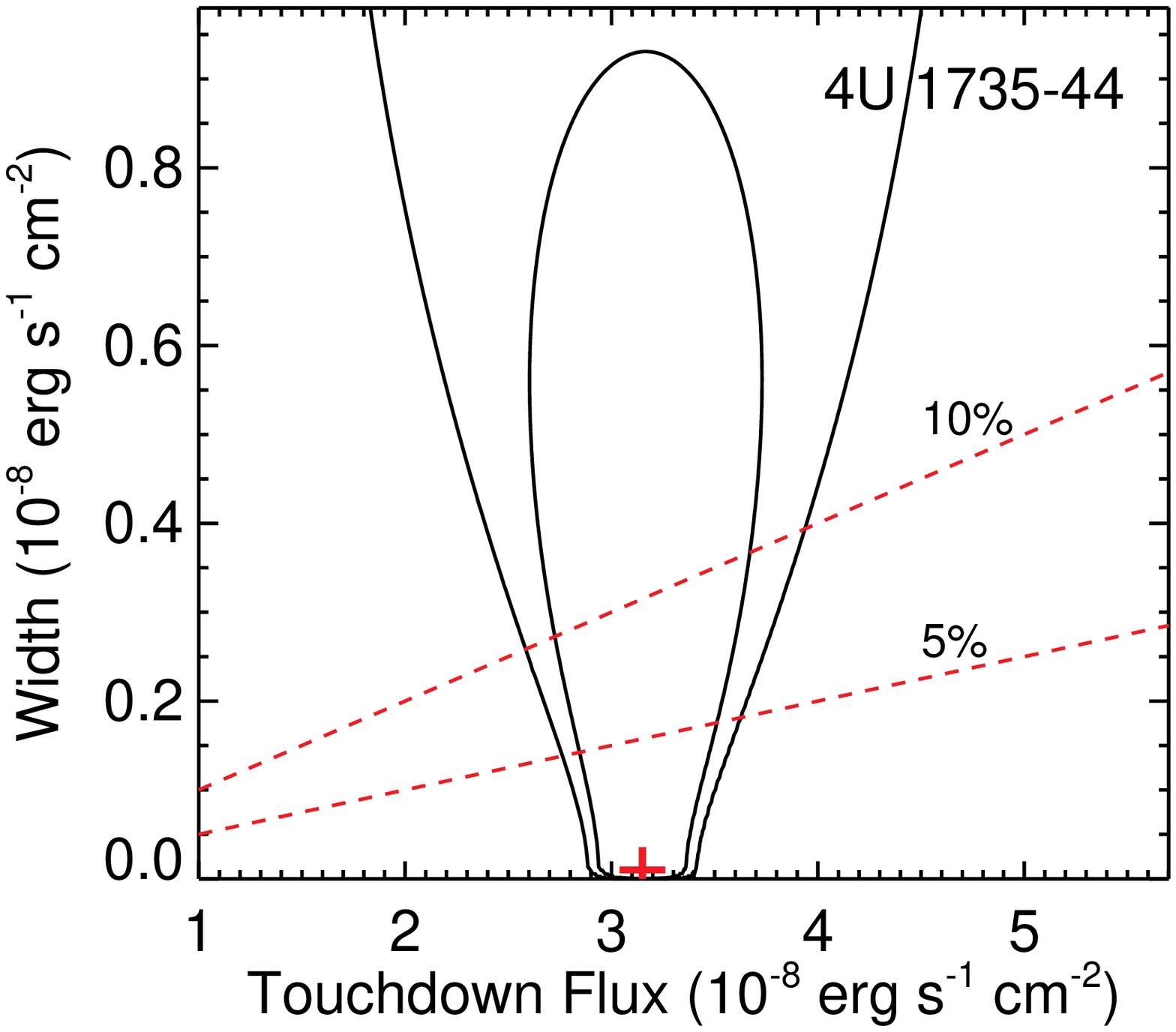} 
   \caption{Same as Figure~\ref{tdex} for the sources 4U~0513$-$401, 4U~1724$-$307,
   KS~1731$-$26, and 4U~1735$-$44.}
\label{td1}
\end{figure}

\begin{figure}
\centering
    \includegraphics[scale=0.3, angle=0]{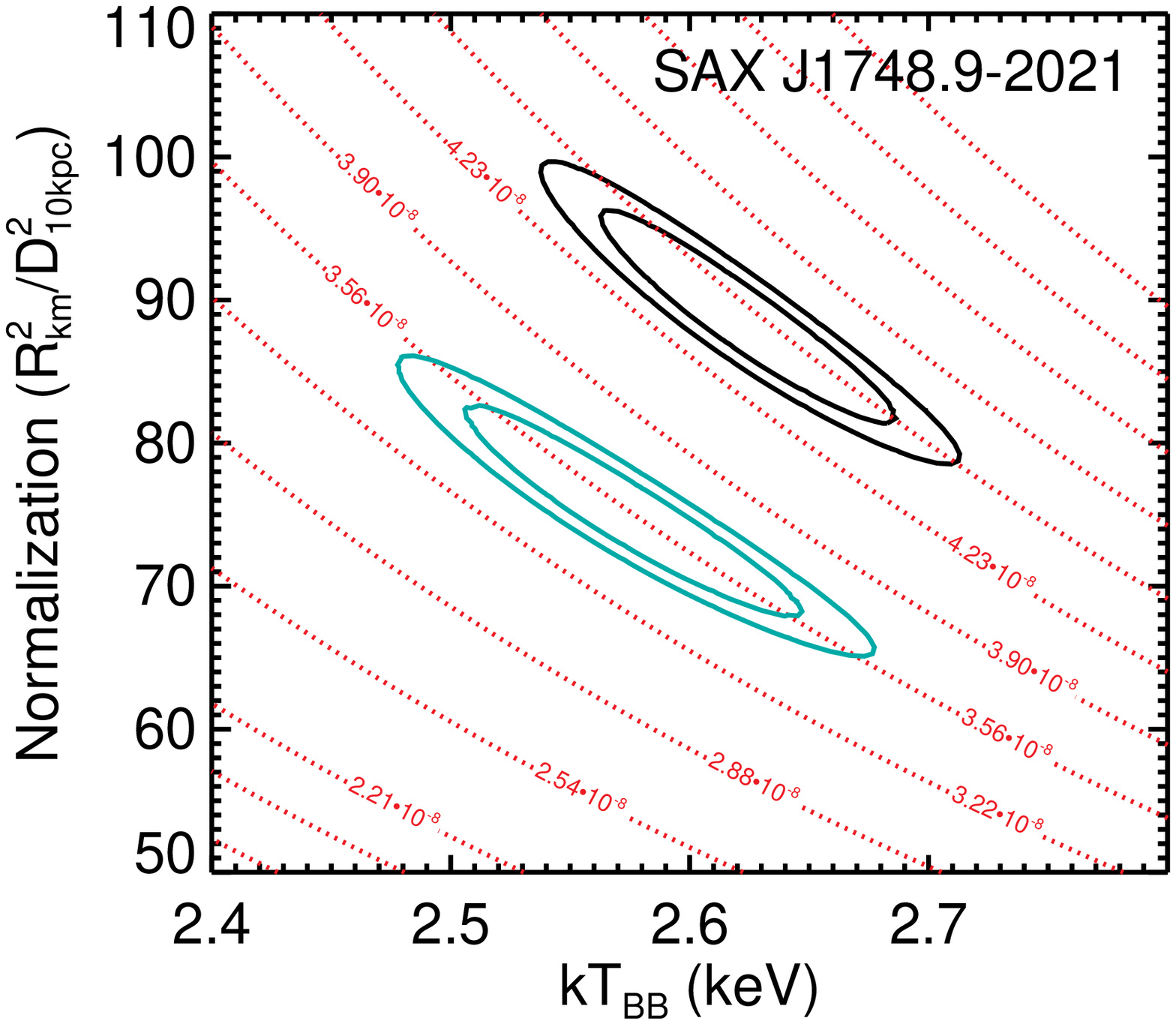}
   \includegraphics[scale=0.3, angle=0]{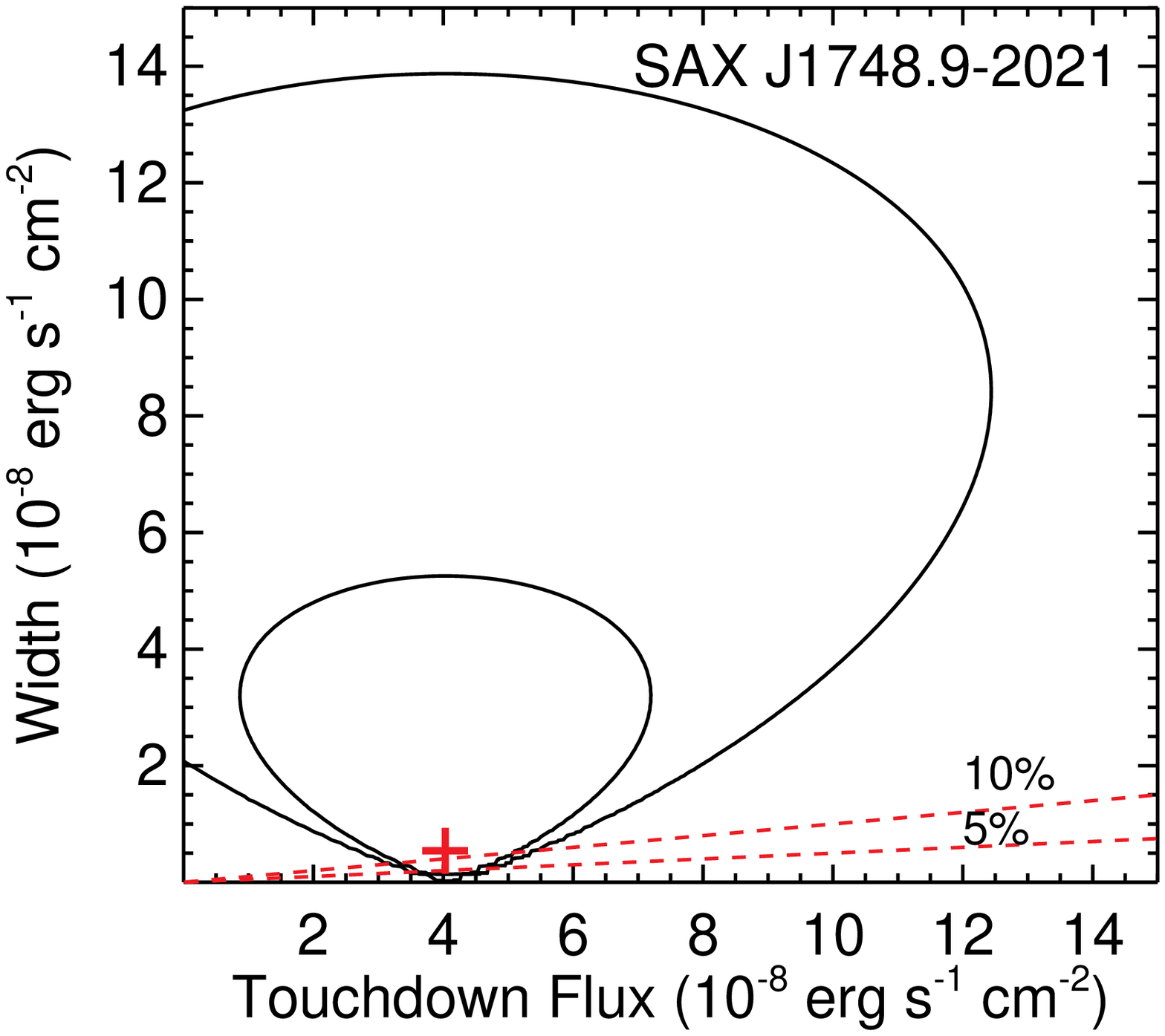}
    \includegraphics[scale=0.3, angle=0]{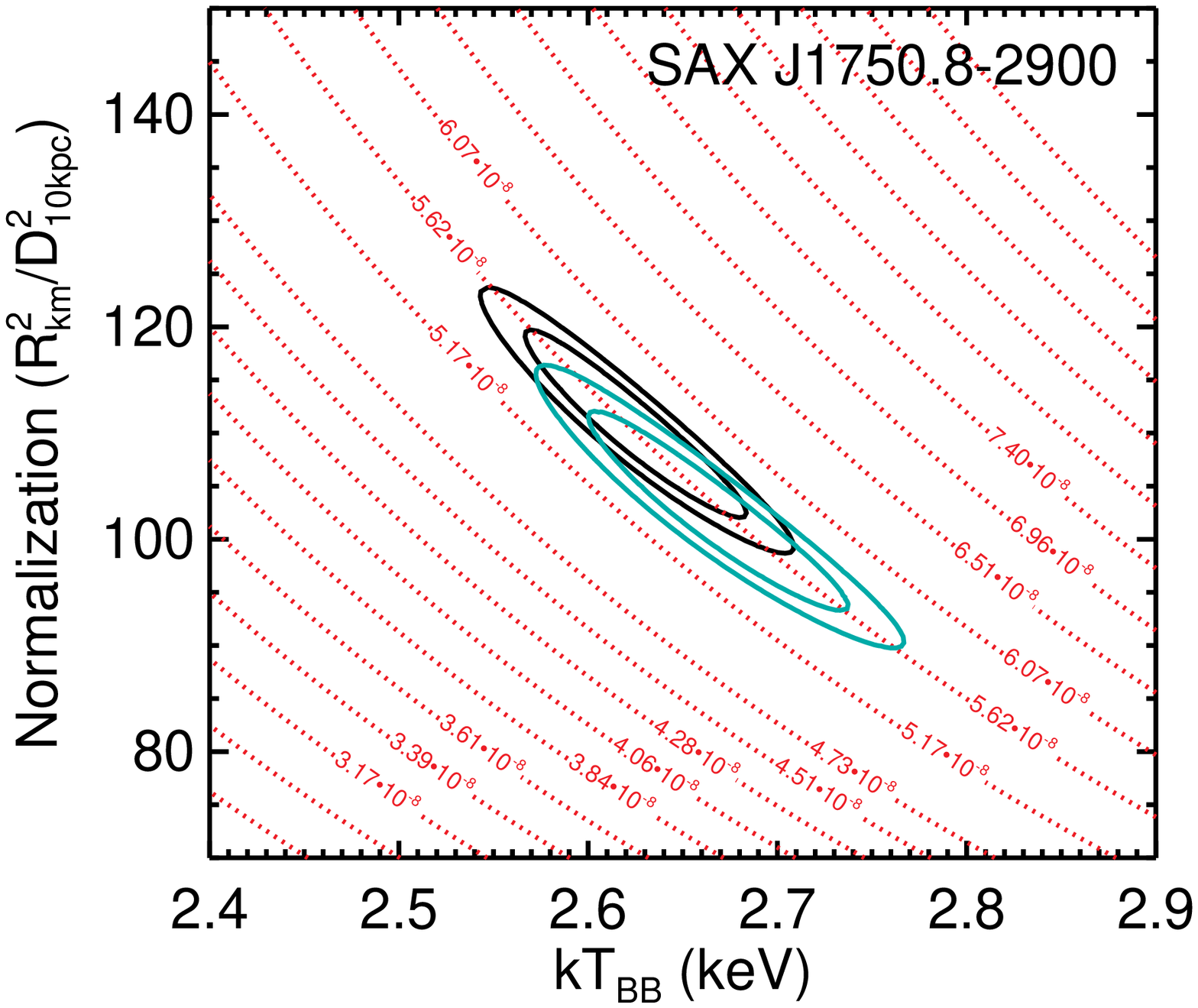}
    \includegraphics[scale=0.3, angle=0]{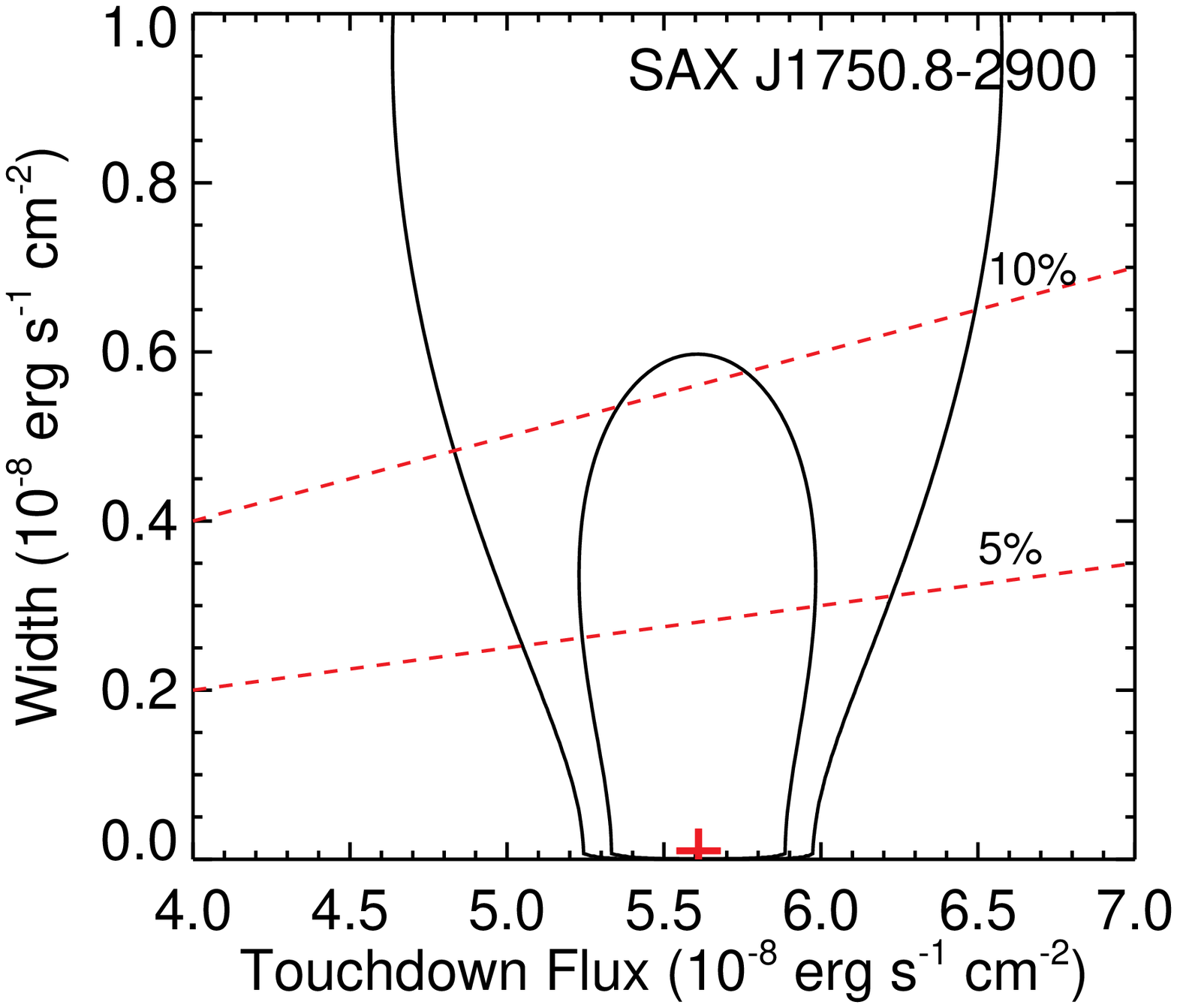}
    \includegraphics[scale=0.3, angle=0]{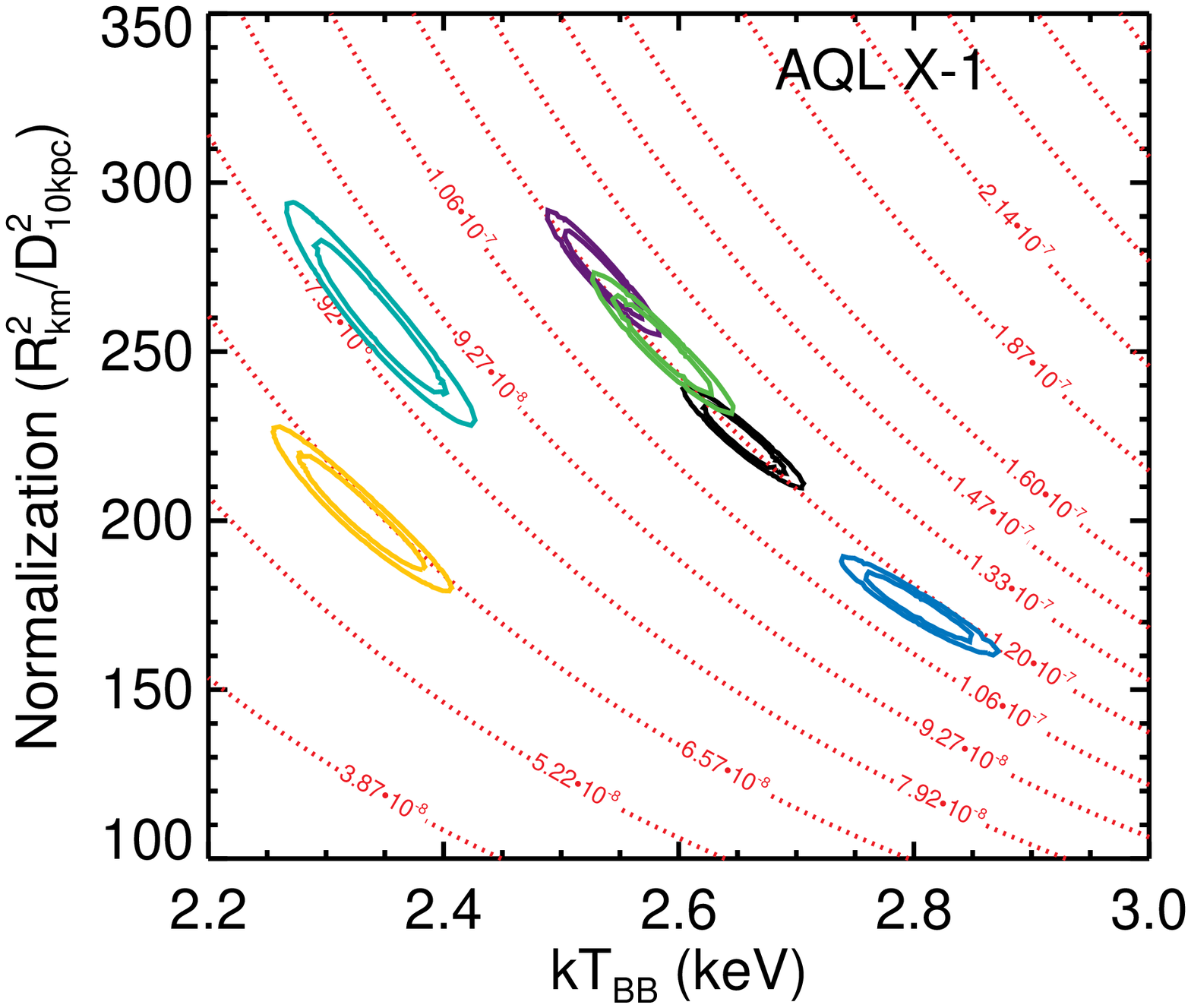}
    \includegraphics[scale=0.3, angle=0]{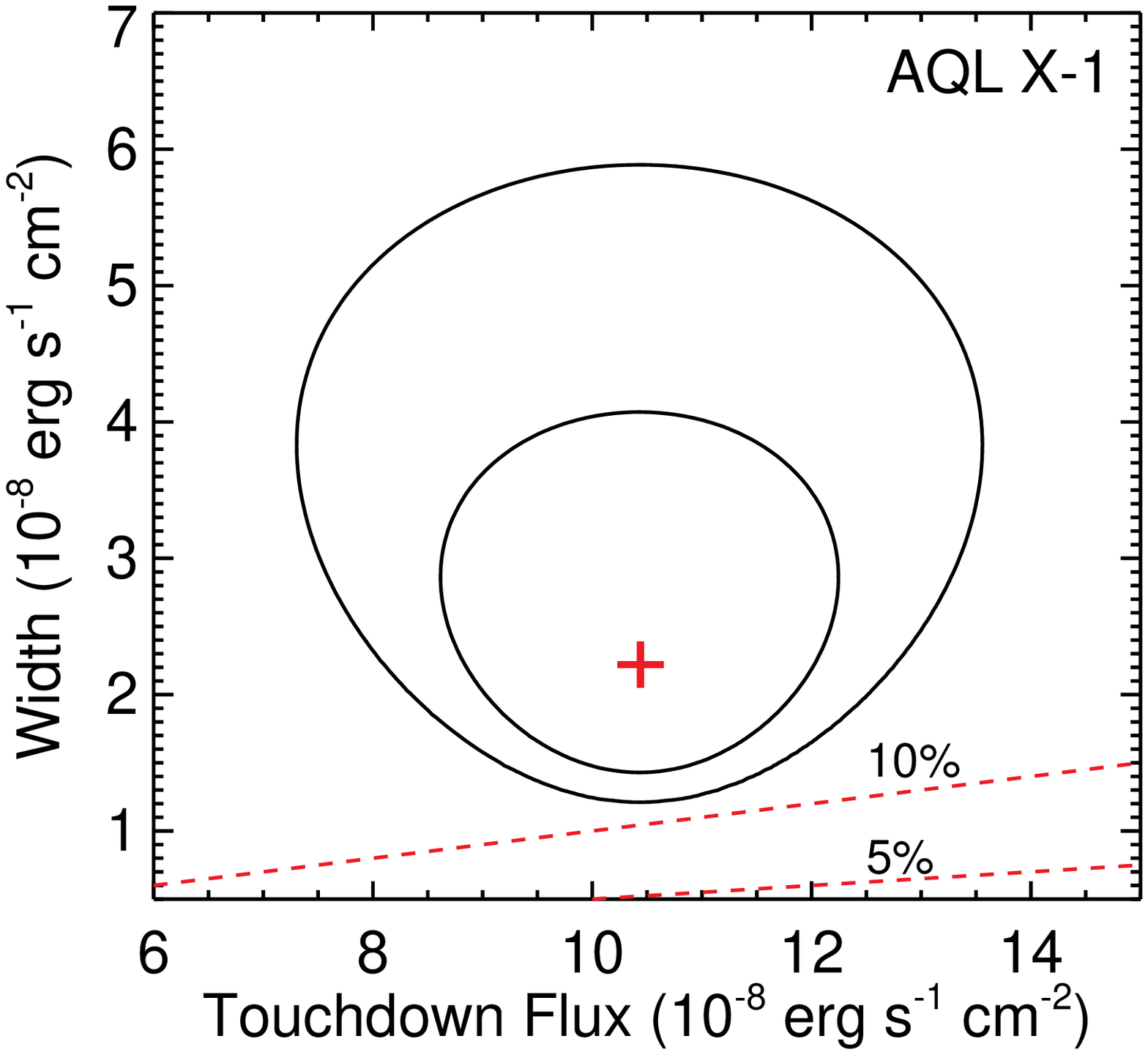}
    \caption{Same as Figure~\ref{td1}, for the sources SAX~J1750.8$-$2900,
      SAX~J1748.9$-$2021, and Aql~X$-$1.}
\label{td2}
\end{figure}

Even though  it is difficult to  determine the shape and  width of the
underlying distribution for any of the sources with only 2 PRE bursts,
it  is worth noting  that for  3 out  of the  6 cases,  the fractional
difference  between the  two  touchdown fluxes,  $F_1$  and $F_2$,  as
defined by

\begin{equation}
R\equiv \frac{2(F_1 - F_2)}{F_1+F_2}\;,
\label{eq:R}
\end{equation}
is  less than  7\%.   It would  be  very unlikely  for the  underlying
distribution of  touchdown fluxes  in each source  to be  much broader
than this level and for half of the randomly picked pairs of touchdown
fluxes to be within 7\%.

In the  following section,  we quantify this  statement by  making the
assumption that  all sources have  a distribution of  touchdown fluxes
with the same  fractional width. We use the $R$ value  for each of the
burst pairs given in Table~\ref{table:R}  to show that the most likely
fractional width of the underlying distribution of touchdown fluxes is
$11^{+5}_{-3}\%$ (68\% confidence level).

\section{Systematic Uncertainties in Sources with few PRE Bursts}

Our aim here  is to estimate the most  likely fractional dispersion of
touchdown fluxes in X-ray bursters that can reproduce the observed $R$
values for  the six sources in our  sample for which we  have only two
observations of PRE  bursts each. Because of the  small number of data
points available,  we will assume that the  underlying distribution of
touchdown  fluxes  in  each  source  is  a  Gaussian,  with  the  same
fractional dispersion $\sigma$, i.e.,

\begin{equation}
P_{\rm td}(F/F_0;\sigma) = \frac{1}{\sqrt{2 \pi \sigma^2}} 
\exp\left[-\frac{(F/F_0 -1)^2}{2\sigma^2} \right]\;,
\end{equation}
where $F$ is the touchdown flux of each burst and $F_0$ is the mean
touchdown flux for each source.

If we draw a random pair of touchdown fluxes $F_1$ and $F_2$ from this
distribution   and   calculate   their   fractional   difference   $R$
(eq.~[\ref{eq:R}]),  then the  distribution of  the R  values  will be
given by
\begin{equation}
P(R;\sigma) = C \int P_{\rm td}\left(\frac{F}{F_0};\sigma\right)
\left\{P_{\rm td}\left[\left(\frac{2-R}{2+R}\right)\frac{F}{F_0};\sigma\right]
+ P_{\rm td}\left[\left(\frac{2+R}{2-R}\right)\frac{F}{F_0};\sigma\right] 
\right\} d\left(\frac{F}{F_0} \right)\;,
\end{equation}
where $C$  is an appropriate normalization  constant. The distribution
$P(R;\sigma)$  peaks at  $R=0$ for  all values  of $\sigma$  and drops
quickly  to  zero  such  that   the  median  value  of  $R$  for  this
distribution  is  $R_{\rm 50\%}=\sigma$.  Given  that  half  of our  6
sources with  only pairs of PRE  bursts have $R$ values  that are less
than 7\%,  we expect  that the most  probable value of  the fractional
dispersion of their touchdown fluxes will be of the same order.

\begin{figure}
\centering
   \includegraphics[scale=0.7, angle=0]{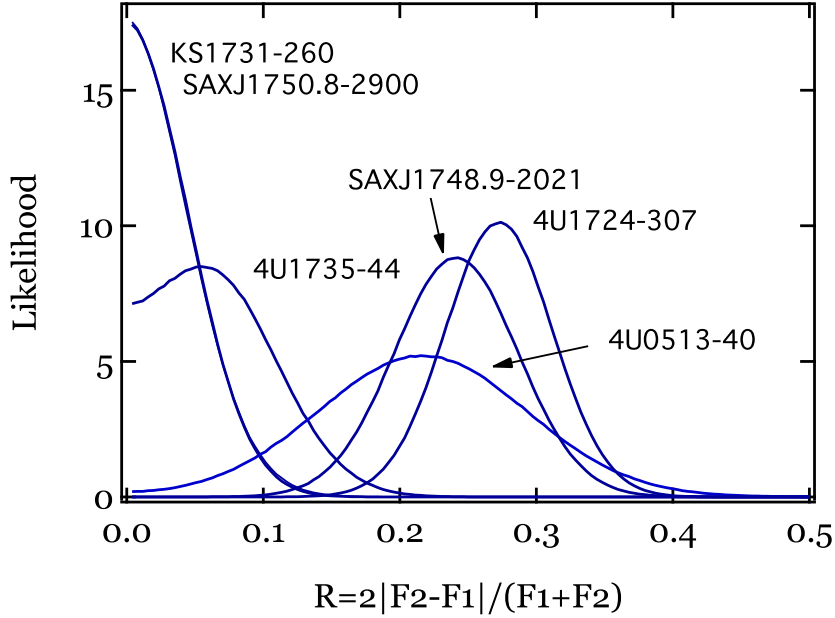}
   \caption{The likelihood of the fractional difference $R$ between
     the touchdown fluxes, $F_1$ and $F_2$, of pairs of bursts in
     sources with only two PRE bursts.}
\label{Rvalue}
\end{figure}

For  each source  with a  pair  of PRE  bursts, we  assign a  Gaussian
likelihood of $R$ values, taking  into consideration the fact that the
$R$ value is always positive, as
\begin{equation}
P_{\rm obs}(R;R_0^i, \sigma_R^i)=\frac{1}{\sqrt{2\pi}\sigma_R^{i}}
\left\{
\exp\left[-\frac{(R-R_0^i)^2}{2(\sigma_R^{i})^2}\right]+
\exp\left[-\frac{(-R-R_0^i)^2}{2(\sigma_R^{i})^2}\right]\right\}\;,\qquad
R>0\;.
\end{equation}
with a most likely value $R_0^i$ and a dispersion $\sigma_R^i$ given
in Table~\ref{table:R}. The likelihood $P_{\rm obs}(R;R_0^i;\sigma_R^i)$ 
for each source with a pair of PRE bursts is shown in Figure~\ref{Rvalue}.

The likelihood of observing the $N=6$ pairs of $R$ values with the
likelihood shown in Figure~\ref{Rvalue}, given an underlying fractional
dispersion $\sigma$, is
\begin{equation}
P({\rm data}\vert \sigma) = \prod_{i=1}^{N} \int P(R;\sigma) 
P_{\rm obs}(R;R_0^i, \sigma_R^i) dR\;.
\end{equation}
Using Bayes' theorem, we can then calculate the likelihood of
each fractional dispersion $\sigma$ given the data, as
\begin{equation}
P(\sigma \vert {\rm data}) = C^\prime P({\rm data} \vert \sigma)
P_\sigma(\sigma)\;.
\end{equation}
where $C^\prime$ is another appropriate normalization constant and we
take the prior probability over all possible fractional dispersions
$P_\sigma(\sigma)$ to be constant over the range of interest.

\begin{figure}
\centering
   \includegraphics[scale=0.7, angle=0]{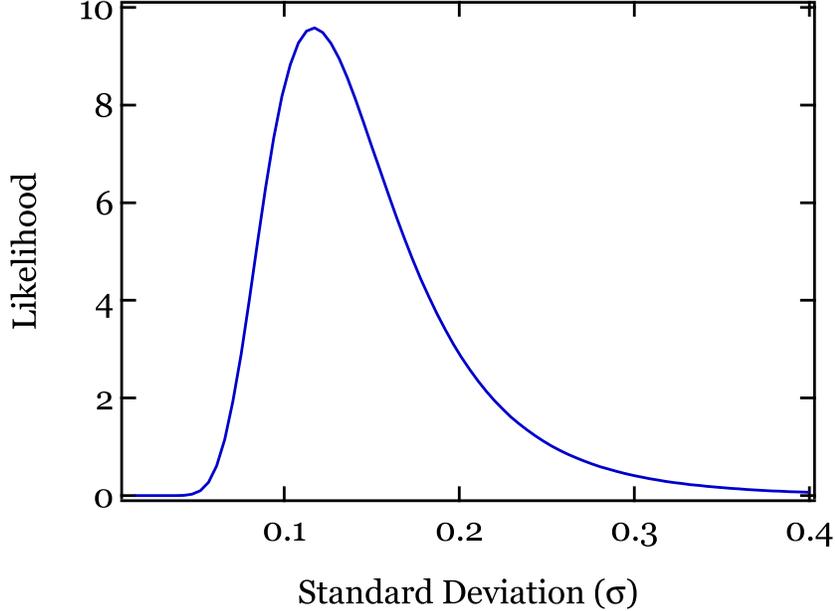}
   \caption{The posterior probability over the fractional width
     $\sigma$ of the touchdown flux distribution for the six sources
     that exhibit only a pair of PRE bursts each.}
\label{sigma_prob}
\end{figure}

Figure~\ref{sigma_prob}  shows  the  posterior  probability  over  the
fractional dispersion $\sigma$ that  is consistent with the 6 observed
$R$ values. The most  likely fractional dispersion of touchdown fluxes
for our sample of 6 sources  with only one pair of observed PRE bursts
each is  $11^{+5}_{-3}$\%, where we determined  the quoted uncertainty
at the 68\% level in the asymmetric probability distribution.

\section{Discussion}

We used the RXTE archive of thermonuclear X-ray bursts to select the
bursts that show a clear evidence for photospheric radius
expansion. We determined systematically the touchdown moment of each
burst and inferred the bolometric flux at that point in the burst.  We
then used a Bayesian technique to infer the most probable value and
the width of the distribution of touchdown fluxes in each source.  In
the two sources with more than a few bursts, the inferred width is
within 5\%$-$10\% of the most probable touchdown flux.  In the six
sources with only one pair of PRE bursts each, the width of the
underlying distribution is consistent with being at a similar
level. When the latter group of sources is taken as a representative
sample, the most likely fractional width of their touchdown fluxes is
$\simeq 11$\%. The only clear exception is Aql X-1, where the
systematic uncertainties exceed $\sim 20\%$.

As we explored in Section~5, the distribution of the touchdown fluxes
is expected to have a finite width for a number of observational and
physical reasons. For a number of these effects, we were able to
estimate that they introduce a 5\%$-$10\% level of systematic
uncertainty in the fluxes. The two unknowns that potentially introduce
larger systematic uncertainties are the asymmetry of the PRE event and
the composition of the material at the photosphere. Our results show,
however, that even these unknowns do not introduce uncertainties
larger than 10\%. 

The PRE bursts allow us to measure the Eddington limit at the surface
of the neutron star for each source. Determining the Eddington limit
requires an absolute flux measurement, which is affected by the
overall flux calibration of the X-ray detector used. Such calibrations
are notoriously difficult to achieve and are usually based on a
particular set of assumptions regarding the spectrum and variability
of the Crab nebula (Jahoda et al.\ 2006; see also Toor \& Seward 1974;
Kirsch et al.\ 2005; Weisskopf et al.\ 2010). Any bias in the absolute
flux calibration cannot increase the spread of touchdown fluxes that
we infer here for each source. However, it can affect the mean
touchdown flux, which, in turn, enters into the measurement of neutron
star masses and radii. We will quantify the potential systematic
uncertainties introduced by the absolute flux calibration of PCA in
Paper III of this series.

It is also important to emphasize here that our results are based on a
statistical analysis of the entire sample of PRE bursts per source and
do not preclude the possibility that any one individual burst may show
a rather different touchdown flux. Indeed, there is at least one burst
observed from 4U~1636$-$536 (ID \#16)\footnote{Although this burst is
a PRE event and we find a touchdown flux that is very similar to the
one in Galloway et al.\ (2006), it is not included in this study
because the persistent flux of the binary before this X-ray burst was
higher than our limit.}, for which the touchdown flux was smaller
compared to the average value by a factor of 1.7~Galloway et al.\
(2006).  In this particular case, a variation in the hydrogen mass
fraction from $X=0$ to $X=0.7$ between the bursts has been considered
as a natural explanation of the difference in touchdown fluxes
(Sugimoto et al.\ 1984; Galloway et al.\ 2006).  The fact that such
outliers may and do exist makes it essential that proper statistical
tools are used in all inferences based on measurements of the
touchdown fluxes of PRE bursts.

In conclusion, our results indicate that the systematic uncertainties
in the measurements of touchdown fluxes in radius-expansion bursts
from low-mass X-ray binaries are within $\simeq 10$\%, for nearly the
entire source sample. Such systematic uncertainties do not preclude,
in and of themselves, neutron star mass-radius measurements with high
enough precision to distinguish between different equations of state
of neutron-star matter.

\LongTables
\begin{deluxetable}{lcccc}
\tablecolumns{5}  
\tablewidth{0pc}  
  \tablecaption{Measured Touchdown Flux Values From PRE Events.}
\tablehead{  \colhead{}   &    \colhead{}   &   \colhead{} & \colhead{Touchdown} 
      &      \colhead{Normalization}
  \\
  \colhead{Source   Name}   &    \colhead{BID\tablenotemark{a}}   &  
\colhead{MJD\tablenotemark{a}}  & 
   \colhead{Flux\tablenotemark{b}}&  \colhead{Ratio} }
\startdata
4U~0513$-$40       & 6 &53442.08752& 1.32\er0.07 & 3.30 \\*
                   & 7 &54043.68856& 1.06\er0.06 & 9.00 \\
\hline
4U~1636$-$53  &1 &  50445.94404  &     7.25\er0.15 & 6.63 \\
              &4 &  50448.73395  &     7.09\er0.15 & 6.19 \\
   	      &6 &  51044.48934  &     7.43\er0.19 & 4.95\\
  	      &7 &  51045.15288  &     7.64\er0.23 & 9.82 \\
 	      &10&  51297.07198  &     7.55\er0.21 & 8.21 \\
 	      &12&  51339.24688  &     7.23\er0.18 & 4.82 \\
              &13&  51347.98824  &     6.35\er0.16 & 4.16 \\
 	      &14&  51348.72984  &     6.86\er0.15 & 5.01 \\
 	      &15&  51350.79575  &     6.52\er0.14 & 5.12 \\
 	      &20&  51710.21233  &     7.81\er0.20 & 7.24 \\
 	      &21&  51765.05463  &     6.28\er0.20 & 5.71 \\
 	      &22&  51765.37284  &     7.00\er0.40 & 5.60 \\
 	      &23&  51768.98081  &     7.52\er0.18 & 5.90 \\
 	      &24&  51820.98112  &     7.24\er0.17 & 5.47 \\
 	      &25&  51853.18194  &     6.64\er0.22 & 3.07 \\
 	      &26&  51860.75171  &     6.02\er0.16 & 3.74 \\
 	      &27&  51937.11612  &     6.70\er0.16 & 5.65 \\
 	      &28&  51941.87558  &     6.43\er0.16 & 3.74 \\
 	      &29&  51942.10024  &     6.62\er0.23 & 5.26 \\
 	      &30&  52004.71326  &     6.65\er0.17 & 7.54 \\
              &31&  52029.22818  &     6.74\er0.16 & 4.04 \\
 	      &34&  52075.13477  &     7.97\er0.25 & 9.89 \\
 	      &38&  52149.27871  &     6.35\er0.15 & 2.67 \\
  	      &45&  52182.61618  &     8.11\er0.22 & 5.10 \\
 	      &49&  52283.01850  &     6.93\er0.18 & 4.36 \\
              &50&  52273.69081  &     5.56\er0.17 & 7.02 \\
 	      &61&  52286.05404  &     8.36\er0.20 & 5.01 \\
 	      &62&  52286.55466  &     7.42\er0.20 & 6.78 \\
 	      &68&  52287.52190  &     6.15\er0.33 & 8.32 \\
      	      &72&  52288.51431  &     6.85\er0.20 & 6.28 \\
 	      &79&  52288.97438  &     5.60\er0.17 & 6.52 \\
 	      &86&  52289.29282  &     7.89\er0.21 & 5.92 \\
 	      &87&  52289.97694  &     6.43\er0.20 & 8.64 \\
 	      &88&  52304.96314  &     5.84\er0.16 & 4.59 \\
 	      &94&  52310.93185  &     6.69\er0.17 & 7.21 \\
 	      &110& 52316.73272  &     7.06\er0.20 & 4.24 \\
  	      &111& 53516.31312  &     7.05\er0.17 & 4.47 \\
	      &122& 52551.25121  &     6.26\er0.17 & 1.89 \\
 	      &136& 53516.31312  &     7.81\er0.24 & 10.17\\
              &137& 53524.38883  &     7.70\er0.20 & 7.47 \\
 	      &148& 53592.23376  &     7.33\er0.17 & 5.69 \\
              &149& 53596.08782  &     6.37\er0.22 & 4.62 \\
  	      &150& 53598.07334  &     7.12\er0.26 & 6.37 \\
  	      &168& 53688.95192  &     7.36\er0.20 & 7.32 \\
\hline
4U~1702$-$429 & 19 & 52957.62907 & 9.05\er0.26 & 3.12 \\
\hline
4U~1705$-$44& 5 & 50542.50287 & 4.13\er0.13 & 2.63 \\
\hline
4U~1724$-$307 & 2 & 53058.40140 & 4.56\er0.13 & 1.76 \\*
              & 3 & 53147.21828 & 6.01\er0.17 & 1.67 \\
\hline
4U~1728$-$34  & 2&  50128.88220 & 8.13\er0.17 & 2.30 \\*
	      & 21& 50718.47163 & 9.21\er0.27 & 3.04 \\*
	      & 22& 50718.66257 & 8.41\er0.16 & 4.47 \\*
	      & 38& 51133.42394 & 8.88\er0.23 & 3.32 \\*
	      & 39& 51133.67299 & 8.36\er0.21 & 2.46 \\*
	      & 41& 51134.57233 & 8.97\er0.23 & 2.35 \\*
              & 48& 51204.00117 & 8.50\er0.19 & 2.65 \\*
              & 49& 51204.12990 & 8.86\er0.28 & 1.80 \\*
              & 51& 51206.14068 & 8.86\er0.19 & 4.05 \\*
              & 53& 51209.91806 & 8.16\er0.26 & 1.86 \\*
              & 54& 51210.08245 & 8.18\er0.18 & 1.85 \\*
              & 55& 51213.93849 & 8.80\er0.19 & 2.04 \\*
              & 69& 51443.01361 & 8.43\er0.24 & 1.66 \\*
              & 83& 51949.12600 & 10.68\er0.38 & 1.99 \\*
              & 85& 52007.61313 & 8.09\er0.20 & 2.03 \\*
              & 86& 52008.08709 & 8.29\er0.20 & 3.38 \\
\hline
KS~1731$-$260 & 8 & 51235.71747 &4.65\er0.13 & 4.49 \\*
              & 9 & 51236.72580 &4.75\er0.13 & 3.90 \\
\hline
4U~1735$-$44  & 6 & 50963.42981 &3.27\er0.12 & 2.68 \\*
              & 7 & 50963.48944 &3.07\er0.10 & 2.26 \\
\hline
SAX~J1748.9$-$2021  & 1& 52190.38947 & 4.52\er0.14 & 33.99 \\*
                    & 2& 52190.46882& 3.54\er0.12 & 3.37 \\
\hline
SAX~J1750.8$-$2900  & 2& 52011.59758 & 5.63\er0.16 & 1.86 \\*
		    & 3& 52014.71002 & 5.58\er0.19 & 2.13 \\
\hline
Aql~X$-$1  & 4 & 50508.97681 & 11.95\er0.19 & 3.91 \\*
           & 5 & 50696.52359 & 12.16\er0.19 & 5.56 \\*
           & 10& 51332.77990 & 11.55\er0.24  & 7.52 \\*
           & 19& 51856.15690 & 8.45\er0.25  & 7.36 \\*
           & 28& 52324.99055 & 12.09\er0.21 & 6.22 \\*
           & 29& 52347.18234 & 6.38\er0.18  & 2.24 
\enddata
\tablenotetext{a}{Burst  IDs  and burst  start  times  are  adopted from (Galloway et al. (2008).}  
\tablenotetext{b}{Values are  given  in  units  of  10$^{-8}$  ergs  cm$^{-2}$  s$^{-1}$  and  are
 calculated using the equation 3 of Galloway et al. (2008).}
\label{tdres}
\end{deluxetable}

\begin{deluxetable}{lccc}
 \tablecolumns{4}   
\tablewidth{300pt}
\tablecaption{The Measured Touchdown Fluxes} 
\tablehead{
\colhead{Source Name}   &  \colhead{Touchdown Flux\tablenotemark{a}} & 
\colhead{$\sigma_{Sys}$\tablenotemark{b}}      
& \colhead{$\sigma_{Formal}$\tablenotemark{c}}}
  \startdata
  4U~0513$-$401       & 1.19 & 0.11 & 0.06\\
  4U~1636$-$536       & 6.93 & 0.64 & 0.20\\
  4U~1724$-$307       & 5.29 & 0.70 & 0.16 \\
  4U~1728$-$34        & 8.63 & 0.46 & 0.22 \\
  KS~1731$-$260       & 4.71 & n/a & 0.13 \\
  4U~1735$-$44        & 3.15 & n/a & 0.11 \\
  SAX~J1748.9$-$2021  & 4.03 & 0.54 & 0.13 \\
  SAX~J1750.8$-$2900  & 5.61 & 0.01 & 0.17 \\
  Aql~X$-$1           & 10.44 & 2.22 & 0.21
\enddata
\tablenotetext{a}{Fluxes and uncertainties are in units  of  10$^{-8}$ erg  s$^{-1}$cm$^{-2}$}
\tablenotetext{b}{These reflect the most probable widths of the underlying distributions}
\tablenotetext{c}{These reflect the uncertainties in measuring the most probable values 
of the underlying distributions}
\label{tdaverage}
\end{deluxetable}

\newpage
\begin{deluxetable}{lc}
 \tablecolumns{2}   
\tablewidth{200pt}
\tablecaption{Fractional Differences of Pairs of Touchdown Fluxes}
\tablehead{
\colhead{Source Name}   &  \colhead{$R$\tablenotemark{a}}}
  \startdata
  4U~0513$-$401       & 0.218$\pm$0.077\\
  4U~1724$-$307       & 0.274$\pm$0.039\\
  KS~1731$-$260       & 0.021$\pm$0.039\\
  4U~1735$-$44        & 0.063$\pm$0.049\\
  SAX~J1748.9$-$2021  & 0.243$\pm$0.045\\
  SAX~J1750.8$-$2900  & 0.009$\pm$0.044
\enddata
\tablenotetext{a}{defined as $R\equiv 2\vert F_2-F_1\vert/(F1+F2)$}
\label{table:R}
\end{deluxetable}

\acknowledgements 

We thank the anonymous referee for constructive suggestions. This work
was supported by NASA ADAP grant NNX10AE89G and Chandra Theory grant
TMO-11003X. DP was supported by the NSF CAREER award NSF 0746549 and
Chandra Theory grant TMO-11003X.  This research has made use of data
obtained from the High Energy Astrophysics Science Archive Research
Center (HEASARC), provided by NASA's Goddard Space Flight Center.

\end{document}